\documentclass[12pt,a4paper]{article}
\usepackage{amsmath,amssymb,hyperref,bm}
\usepackage{graphicx}
\usepackage{cite}
\numberwithin{equation}{section}
\allowdisplaybreaks[3]

\newcommand{\cO}{\mathcal{O}}
\newcommand{\cN}{\mathcal{N}}

\newcommand{\bb}{\bm{b}}

\begin{document}
%%% Title page %%%%%
\begin{titlepage}

\renewcommand{\thefootnote}{\fnsymbol{footnote}}
\begin{flushright}
\begin{tabular}{l}
YITP-18-06\\
%\today %This should be commented out.
\end{tabular}
\end{flushright}

\vfill
\begin{center}

% \vskip 2.5 truecm

\noindent{\large \textbf{Conformal blocks from Wilson lines with loop corrections}}

%\medskip
%
%\noindent{\large \textbf{} }}

\vspace{1.5cm}

\noindent{Yasuaki Hikida$^{a}$\footnote{E-mail: yhikida@yukawa.kyoto-u.ac.jp} and Takahiro Uetoko$^b$\footnote{E-mail: rp0019fr@ed.ritsumei.ac.jp}}

\bigskip

\vskip .6 truecm

\centerline{\it $^a$Center for Gravitational Physics, Yukawa Institute for Theoretical Physics,}
\centerline{\it  Kyoto University, Kyoto 606-8502, Japan}
\medskip
\centerline{\it $^b$Department of Physical Sciences, College of Science and Engineering,}
\centerline{\it Ritsumeikan University, Shiga 525-8577, Japan}

\end{center}

\vfill
\vskip 0.5 truecm

\begin{abstract}

We compute the conformal blocks of the Virasoro minimal model or its W$_N$ extension with large central charge from Wilson line networks in a Chern-Simons theory including loop corrections. In our previous work, we offered a prescription to regularize divergences from loops attached to Wilson lines. In this paper, we generalize our method with the prescription by dealing with more general operators for $N=3$ and apply it to the identity W$_3$ block. We further compute general light-light blocks and heavy-light correlators for $N=2$ with the Wilson line method and compare the results with known ones obtained using a different prescription.
We briefly discuss general W$_3$ blocks.

\end{abstract}
\vfill
\vskip 0.5 truecm

\setcounter{footnote}{0}
\renewcommand{\thefootnote}{\arabic{footnote}}
\end{titlepage}

\newpage

\tableofcontents
%%%%%%%%%%%%%%%%%%%%%%%%%%%%%%%%%%%%%%%%%%%%%%%%%%%%%%%%%%%%%%%%%%%%%%

\section{Introduction}

Two dimensional conformal field theory (CFT) with large central charge $c$ is supposed to be dual to the semi-classical regime of three dimensional gravity on anti-de Sitter (AdS) space. Since holography is expected to provide a formulation of quantum gravity in terms of dual CFT, we would obtain some insights on the quantum aspects of gravity from $1/c$ corrections in dual CFT.
As a simple example, we examine the W$_N$ minimal model, which has a coset description as
\begin{align}
\frac{\text{su}(N)_k \oplus \text{su}(N)_1 }{\text{su}(N)_{k+1}} 
\label{coset}
\end{align}
with the central charge
\begin{align}
c = (N-1) \left( 1 - \frac{N(N+1)}{(k+N)(k+N+1)}\right) \, .
\label{center}
\end{align}
The dual bulk description was proposed to be the Chern-Simons gauge theory based on  $\text{sl}(N) \oplus \text{sl}(N)$ Lie algebra \cite{Gaberdiel:2010pz}. In particular, we can study the quantum corrections of AdS$_3$ gravity in terms of the Virasoro minimal model given by the coset \eqref{coset} with $N=2$.
We consider the large $c$ regime of the minimal model, which corresponds to a perturbative regime of Chern-Simons theory. 
The expression of central charge \eqref{center} implies that $k = - 1 - N + \mathcal{O}(c^{-1})$; thus, 
the large $c$ regime of the model should be non-unitary.
Nevertheless, the holography has been argued to work
 \cite{Castro:2011iw,Gaberdiel:2012ku,Perlmutter:2012ds}, and it should be useful at least for problems related to its symmetry algebra.

The basic quantities of CFT are the spectrum of operators and correlation functions or conformal blocks.
In particular, the conformal block decomposition of the correlator is an essential tool for recent developments on conformal bootstrap programs (see, e.g., \cite{ElShowk:2012ht}).
In terms of Chern-Simons theory, conformal blocks at the large $c$ limit were proposed to be computed by the networks of open Wilson lines, and the proposal was confirmed for explicit examples \cite{Bhatta:2016hpz,Besken:2016ooo}.
In this paper, we develop techniques to evaluate conformal blocks beyond the large $c$ limit by applying recent works on loop effects for Wilson lines in  \cite{Fitzpatrick:2016mtp,Besken:2017fsj,Hikida:2017ehf}, see also \cite{Anand:2017dav}.
The proposal of the map between conformal blocks and Wilson line networks in \cite{Bhatta:2016hpz,Besken:2016ooo} can be easily extended  beyond the large $c$ limit as in \cite{Besken:2017fsj}. We evaluate the $1/c$ corrections of conformal blocks from the loop corrections of the Wilson line networks.
The computations are rather subtle,
and a main issue is how to regularize divergences associated with loop diagrams.
In the language of the usual Lagrangian formulation, how to deal with this kind of divergences was established. Since we are dealing with open Wilson lines instead of Lagrangian, it is quite non-trivial to formulate a general procedure. In our previous paper \cite{Hikida:2017ehf}, we offered a way to regularize divergences by renormalizing parameters in Wilson lines, which correspond to couplings between bulk scalars and a higher spin current, along with rescaling the overall factor of open Wilson lines.
Moreover, we showed that ambiguities related to the renormalization procedure can be removed by requiring that the correlators of boundary theory are consistent with the symmetry.
In this paper, we adopt the renormalization procedure in \cite{Hikida:2017ehf} for regularizing  divergences.

The primary operators of the W$_N$ minimal model \eqref{coset} are labeled as $(\Lambda_+ ; \Lambda_-)$, where $\Lambda_\pm$ represent the highest weights of su$(N)$ Lie algebra.
The operator with $(\text{f};0)$ (where f denotes the fundamental representation of su$(N)$) is dual to a bulk scalar field,
and the operators with $(\Lambda;0)$ correspond to their composite fields \cite{Gaberdiel:2010pz}. On the other hand, the operator with $(0;\Lambda)$ has its conformal weight of order $c$, and it is proposed to be dual to a conical defect geometry \cite{Castro:2011iw,Gaberdiel:2012ku,Perlmutter:2012ds}.
The symmetry of W$_N$ algebra for the model \eqref{coset} is generated by higher spin currents $J^{(s)}$ with $s=2,3,\ldots,N$. In \cite{Hikida:2017ehf}, we evaluated the correlators as
\begin{align}
\langle \bar{\mathcal{O}}_{(\Lambda;0)} (z_1) \mathcal{O}_{(\Lambda;0)} (z_2) \rangle \, , \quad
\langle \bar{\mathcal{O}}_{(\Lambda;0)} (z_1) \mathcal{O}_{(\Lambda;0)} (z_2) J^{(s)} (z_3)\rangle 
\label{23pt}
\end{align}
with general $\Lambda = 2j $ (where $j$ represents the spin of representation) for $N=2$  and with  $\Lambda = \text{f}$ for $N=3$.
In particular, we reproduced the $1/c$ corrections of conformal weight for operators dual to bulk scalars up to $1/c^2$ order by adopting our regularization prescription.%
\footnote{For $N=2$, the conformal weight of operator was reproduced up to $1/c$ order in \cite{Besken:2017fsj} by applying a Wilson line method, which is possible without specifying any regularization prescription.}

In this paper, we extend the previous analysis and apply our Wilson line method to four point conformal blocks with $N=2$ and $N=3$. For $N=2$, we consider the correlator as
\begin{align}
\langle \bar{\mathcal{O}}_{(\Lambda ; 0)} (\infty) \mathcal{O}_{(\Lambda ; 0)} (1)   \bar{\mathcal{O}}_{(\Xi ; 0)} (z) \mathcal{O}_{(\Xi; 0)} (0)  \rangle 
\label{4pt0}
\end{align}
with $(\Lambda , \Xi) = ( 2q , 2 j)$.
The conformal blocks were already examined by utilizing open Wilson lines in \cite{Fitzpatrick:2016mtp}. Since we adopt a different regularization prescription, our formulation is not the same as theirs. 
We can easily relate the two methods for the identity Virasoro block, so we focus on  general Virasoro blocks.%
\footnote{We refer to the conformal block with the identity operator or its descendants exchanged as an identity block and to other conformal blocks as general blocks.} 
In \cite{Fitzpatrick:2016mtp}, they obtained the all-order expression in $z$ for the general Virasoro block up to $1/c$ order except for the $j,q$-independent part. For the part, they obtained the first few terms in $z$ expansion. We compute the general block with our formulation and obtain the all-order expression in $z$ up to $1/c$ order including the $j,q$-independent part.
For $N=3$, we examine the four point conformal block \eqref{4pt0}
with setting $\Xi = \text{f}$ for simplicity.
We explicitly compute the identity W$_3$ block using the product of two open Wilson lines and reproduce the CFT result. For the purpose we first extend the previous analysis for the two and three point functions \eqref{23pt} by dealing with general $\Lambda$  for $N=3$.
The general W$_3$ blocks at the leading order in $1/c$ were obtained in \cite{Fateev:2011qa} from the CFT viewpoint, and they were reproduced in \cite{Besken:2016ooo} using the networks of the open Wilson line. We point out that our formulation combines these results in a quite nice way. 
We also analyze the $1/c$ correction of a simple non-identity W$_3$ block with $(\Lambda , \Xi ) = (\text{f} , \text{f})$.

We further consider the four point function with heavy operators $\mathcal{O}_{(0;\Lambda)}$ as
\begin{align}
\langle \bar{\mathcal{O}}_{(0 ; \Lambda)} (\infty) \mathcal{O}_{(\text{f} ; 0)} (1)   \bar{\mathcal{O}}_{(\text{f} ; 0)} (z) \mathcal{O}_{(0; \Lambda)} (0)  \rangle 
\end{align}
with $N=2$. The heavy operator corresponds to a conical defect geometry as mentioned above, and we compute the correlator using an open Wilson line in the conical space. For quantum corrections to the open Wilson line, we adopt our formulation developed so far. For quantum corrections to the conical geometry, we make use of the results in \cite{Raeymaekers:2014kea} obtained by applying the coadjoint orbits of the Virasoro group in \cite{Witten:1987ty,Alekseev:1988ce}.
In this way, we reproduce the previous results from CFT computation in \cite{Beccaria:2015shq,Fitzpatrick:2015dlt,Chen:2016cms}. 
The Wilson line method was used in \cite{Besken:2016ooo,Fitzpatrick:2016mtp} at the leading order in $1/c$, where they used open Wilson lines for the heavy operators instead of conical spaces.
For general $N$, the computation with our method reduces to those in \cite{Hijano:2013fja,Hegde:2015dqh} at the leading order in $1/c$.

The organization of this paper is as follows.
In the next section, we introduce the W$_N$ minimal model and write down the expressions of conformal blocks examined in terms of Wilson lines in this paper.
In section \ref{BulkMethod}, we explain our bulk method for conformal blocks particularly focusing on our prescription for regularizing divergences.
In subsection \ref{23ptN2}, we review our previous work in \cite{Hikida:2017ehf} on the two and three point functions \eqref{23pt} for $N=2$, and in subsection \ref{WilsonNew}, we extend the analysis by considering general light operators with $N=3$.
In section \ref{CBWilson}, we apply the method to compute conformal blocks from the bulk viewpoint.
In subsection \ref{VacuumCB}, we obtain the identity W$_3$ block from the product of two Wilson lines and compare it to the CFT result.
In subsection \ref{GenericCB}, we examine general Virasoro blocks with our formulation and see the relation to previous results in \cite{Fitzpatrick:2016mtp}. We further discuss general W$_3$ blocks. 
In section \ref{HLcorrelator}, we apply our formulation to correlators with two heavy operators along with two light operators for $N=2$.
We conclude this paper and discuss open problems in section \ref{conclusion}.
In appendix \ref{toda}, we summarize CFT computations of four point conformal blocks in the W$_N$ minimal model.
In particular, we obtain four point function of \eqref{4pt0} with $\Lambda$ corresponding a rectangular Young diagram and $\Xi = \text{f}$.
As an application of  \cite{Hikida:2017byl,Hikida:2017ehf}, we compute the $1/c$ corrections of higher spin charges for the scalar operators using the method of conformal block decomposition.
In appendix \ref{sl3}, we include some technical details of sl$(3)$ Lie algebras.
In appendix \ref{simple}, we compute the $1/c$ correction of a simple non-identity W$_3$ block.
In appendix \ref{Shwarz}, we examine how correlators obtained from Wilson lines change under local conformal transformations.

\section{W$_N$ minimal model}
\label{WNMM}

The minimal model of W$_N$ algebra can be described by  the coset \eqref{coset} with the central charge \eqref{center}, see, e.g., \cite{Bais:1987zk,Bershadsky:1989mf,Bouwknegt:1992wg} for some details.
The primary states are labeled as $(\Lambda_+;\Lambda_-)$ with $\Lambda_\pm$ as the highest weights of  su$(N)$ Lie algebra.
We denote $n^\pm_i$ $(n^\pm_N =0)$ as the numbers of boxes in the $i$-th row of Young diagrams corresponding to $\Lambda^\pm$ and define power sums  as
\begin{align}
C_s (\Lambda_+ ; \Lambda_-) \equiv \frac{2}{ s} \sum_{i=1}^N \left[ \alpha_+ \left (n^+_i  - \frac{|\Lambda_+|}{N} + \rho_i \right) + \alpha_-  \left(n^-_i  - \frac{|\Lambda_-|}{N} + \rho_i \right) \right]^s 
\end{align}
by following \cite{Perlmutter:2012ds}.
Here, $|\Lambda_\pm|$ denote  the numbers of boxes in $\Lambda_\pm$, and $\alpha_\pm$ are given by
\begin{align}
\alpha_+ = \sqrt{\frac{N+k+1}{2(N+k)}} \, , \quad
\alpha_- = - \sqrt{\frac{N+k}{2(N+k+1)}} \, .
\label{alphapm}
\end{align}
The components of the Weyl vector $\bm{\rho}$ are
$
\rho_i = (N+1)/2 - i .
$
The conformal weight of primary state $(\Lambda_+;\Lambda_-)$ is then written as
\begin{align}
h (\Lambda_+;\Lambda_-) = C_2 (\Lambda_+;\Lambda_-) - \frac{c -N +1}{24} \, .
\label{conformalweight}
\end{align}
The two point function of the operator $(\Lambda_ +;\Lambda_-)$ is fixed by the symmetry as
\begin{align}
\label{2pt}
&\langle \bar{\mathcal{O}}_{(\Lambda_+ ;\Lambda_- ) } (z_1)  \mathcal{O}_{(\Lambda_+ ; \Lambda_-) } (z_2)  \rangle
= \frac{1}{z_{12}^{2 h (\Lambda_+; \Lambda_-)}}   \, .
\end{align}
The overall normalization can be set arbitrary by changing the definition of operator. Here and in the following, we only consider the holomorphic sector, but it is straightforward to include the anti-holomorphic sector.

We are interested in a regime with the large central charge \eqref{center} but finite $N$, which implies that we should take
\begin{align}
k = - 1 - N + \frac{N (N^2 -1 )}{c} + \frac{N (1 - N^2) (1 - N^3)}{c^2} + \mathcal{O} (c^{-3}) \, .
\label{kexp}
\end{align}
In the regime, the conformal weight behaves as $\mathcal{O}(c^{0})$ for $\Lambda_- = 0$
while $\mathcal{O}(c)$ for $\Lambda_- \neq 0$. 
For a while, we focus on the case only with light operators $(\Lambda_+ ; \Lambda_-) = (\Lambda ;0)$, and later, we examine correlators with $\Lambda_- \neq 0$ as well.
For  $N=2$, we label the representation of  su(2) by spin quantum number $j =n_1/2$, where $n_i = n_i^+$ is defined above.
The conformal weight of the operator with $j$ is given by
\begin{align}
h_j \equiv h (2j;0)
= - j - \frac{6 j(j+1)}{c}  - \frac{78 j (j+1)}{c^2}  + \mathcal{O}(c^{-3})
\label{hN2}
\end{align}
in $1/c$ expansion.
For general $N$, we expand the highest weight in the fundamental weights $\bm{\omega}_i$ as
$\Lambda = \sum_{i=1}^{N-1} \lambda_i \bm{\omega}_i  $, where $\lambda_i$ are the Dynkin labels.
As a simple but non-trivial example, we also study in some details the $N=3$ case,
where  the Dynkin labels $(\lambda_1,\lambda_2)$ are related as $n_1 = \lambda_1 + \lambda_2$ and $n_2 = \lambda_2$.
The conformal weight of the operator with $(\lambda_1,\lambda_2;0)$ is
\begin{align}
h (\lambda_1 , \lambda_2 ; 0)
&=- \lambda_1 -\lambda_2  -\frac{8 \left(\lambda_1 ^2+\lambda_1  \lambda_2 +3 \lambda_1  +\lambda_2 ^2+3 \lambda_2 \right)}{c}+ \mathcal{O} (c^{-2}) \, . \label{hexp}
\end{align}
For general $N$, the computation of correlators is simplified when $\Lambda = n \bm{\omega}_m$, which corresponds to the case with $n_1 = \cdots = n_m = n$, $n_{m+1} = \cdots n_{N} = 0$.
The conformal weight of the operator with  $\Lambda = n \bm{\omega}_m$ is
\begin{align}
\begin{aligned}
h(n \bm{\omega}_m, 0)
& = \frac{1}{2} m n (m-N) + \frac{m n \left(N^2-1\right) (m-N) (n+N)}{2 c}+\mathcal{O} (c^{-2}) \, .
\end{aligned}
\label{hexpN}
\end{align}
Expanding the conformal weight \eqref{conformalweight} in $1/c$ as
\begin{align}
h (\Lambda ; 0) = h_0 (\Lambda ; 0) + \frac{h_1 (\Lambda ; 0) }{c}  + \mathcal{O}(c^{-2}) \,  ,
\end{align}
the two point function \eqref{2pt} becomes
\begin{align}
\langle \bar{\mathcal{O}}_{(\Lambda ;0 ) } (z_1)  \mathcal{O}_{(\Lambda ; 0) } (z_2)  \rangle
 = \frac{1}{z_{12}^{2 h_0 (\Lambda ; 0) }} - \frac{2 h_1 (\Lambda ; 0)  \log (z_{12})}{c z_{12}^{2 h_0 (\Lambda ; 0) }}  + \mathcal{O}(c^{-2}) \, .
\label{cwshift}
\end{align}
Later we shall see that the bulk computations reproduce the shifts of conformal weight from the terms proportional to $\log(z_{12})$.

The symmetry of W$_N$ algebra is generated by higher spin currents $J^{(s)}(z)$ with $s=2,3,\ldots,N$.
In this paper, we adopt the convention for the higher spin currents in \cite{Hikida:2017byl,Hikida:2017ehf}.
In particular, the two point functions are given by
\begin{align}
\langle J^{(s)} (z_1) J^{(s)} (z_2) \rangle = \frac{B_s}{z_{12}^{2s}}
\label{2ptcurrent}
\end{align}
with
\begin{align}
B_s = \frac{(1 - 2 s ) N_s c }{6} \, , \quad
N_s =  \frac{3 \sqrt{\pi} \Gamma (s) (1 -N)_{s-1} (N+1)_{s-1}}{2^{2s-2} (N^2 - 1) \Gamma(s+\frac12)} \, .
\label{BsNs}
\end{align}
For $s=2,3$, we have
\begin{align}
B_2 = \frac{c}{2} \, , \quad B_3 = \frac{ \left(4 - N^2\right) c}{6}  \, , \quad
N_2 = -1 \, , \quad N_3 = \frac{N^2-4}{5}\, ,
\label{B2B30}
\end{align}
and, in particular, the normalization for the spin 2 current is the same as the usual one for the energy momentum tensor, i.e., $J^{(2)}(z) = T (z)$.
The three point function with a higher spin current $J^{(s)}$ is fixed by the Ward identity as
\begin{align}
\begin{aligned}
&\langle \bar{\mathcal{O}}_{(\Lambda_+ ; \Lambda_-)} (z_1) \mathcal{O}_{(\Lambda_+ ; \Lambda_-)} (z_2)  J^{(s)} (z_3) \rangle \\
& \qquad = w^{(s)} (\Lambda_+ ;\Lambda_-) \left( \frac{z_{12}}{z_{13} z_{23}}  \right)^s
\langle \bar{\mathcal{O}}_{(\Lambda_+ ; \Lambda_-)} (z_1) \mathcal{O}_{(\Lambda_+ ; \Lambda_-)} (z_2)  \rangle \, ,
\end{aligned}
\label{3pt}
\end{align}
where $w^{(s)} (\Lambda_+ ; \Lambda_-) $ denotes the spin $s$ charge of the operator $(\Lambda_+ ; \Lambda_-)$.
The spin 2 charge is nothing but the conformal weight as $w^{(2)} (\Lambda_+ ; \Lambda_-)  = h (\Lambda_+ ; \Lambda_-) $ in \eqref{conformalweight}. The expression of spin 3 charge
$w^{(3)} (\Lambda_+ ; \Lambda_-) \equiv w (\Lambda_+ ; \Lambda_-) $ may be found in \cite{Gaberdiel:2011zw,Castro:2011iw,Perlmutter:2012ds} as
\begin{align}
w (\Lambda_+ ; \Lambda_-) =  \sqrt{\frac{ 2 N (N + 2) \left(1 - N^2\right) }{  \left((c-1) (2 + N)+3 N^2\right)}} C_3 (\Lambda_+ ; \Lambda_-) \, ,
\end{align}
where the overall factor is due to our convention for the spin 3 current.
For $N=3$, the spin 3 charge of the operator $(\lambda_1 , \lambda_2 ; 0)$ is given by
\begin{align}
w(\lambda_1 , \lambda_2 ; 0) = \frac{\lambda_1 - \lambda_2}{3} + \frac{8 (15 \lambda_1+15 \lambda_2+13) (\lambda_1 -\lambda_2)}{15 c} + \mathcal{O} (c^{-2}) \, .
\label{wexp}
\end{align}
For general $N$, the spin 3 charge of the operator $(n \bm{\omega}_m  ; 0)$ is
\begin{align}
\begin{aligned}
& w(n \bm{\omega}_m  ; 0) = \frac{1}{6} m n \left(2 m^2-3 m N+N^2\right) \\ & \qquad +
\frac{m n \left(N^2-1\right) \left(2 m^2-3 m N+N^2\right) (3 n (N+2)+N (N+2)-2)}{6 (N+2) c }+
\mathcal{O}(c^{-2})
\end{aligned}
\label{wexpN}
\end{align}
in $1/c$ expansion.

A main aim of this paper is to examine conformal blocks using open Wilson lines in the bulk theory. 
For $N=2$, we consider the four point function of scalar operators
in \eqref{4pt0} with $(\Lambda,\Xi) = (2 q ,2 j)$, that is
\begin{align}
 \langle \mathcal{O}_{q} (\infty) \mathcal{O}_{q} (1) \mathcal{O}_{j} (z) \mathcal{O}_{j} (0) 
   \rangle \, , 
 \qquad 
 \mathcal{O}_l \equiv \mathcal{O}_{(2 l ; 0)} \, .
 \label{4ptN2}
\end{align}
Here we have used $\bar{\mathcal{O}}_l = \mathcal{O}_l$ since 
the finite representation of su(2) is self-conjugate.
The identity Virasoro block has been evaluated using Wilson lines up to order $1/c^2$  as \cite{Fitzpatrick:2016mtp}
\begin{align}
\begin{aligned}
z^{2 h_j} \mathcal{V}_0 (z) =& 1 + \frac{2 h_j h_q}{c} z^2 {}_2 F_1 (2,2;4;z) \\
& + \frac{1}{c^2} \left[ h_j^2 h_q^2 k_a (z) + (h_j^2 h_q + h_j h_q ^2) k_b (z) + h_j h_q k_c(z) \right] + \mathcal{O}(c^{-3}) \, ,
\end{aligned}
\label{V0}
\end{align}
where  the expressions of $k_a,k_b,k_c$ may be used as in (A.11) of \cite{Hikida:2017byl}. Here, $z^2  {}_2 F_1 (2,2;4;z)$ at the $1/c$ order corresponds to the global block%
\footnote{The global part of Virasoro algebra is given by sl(2) algebra, and only the sl(2) descendants are considered as intermediate states for the global block.}
with the exchange of the energy momentum tensor. 
We do not compute the identity block with the Wilson line method since there is no significant difference from the previous computation in \cite{Fitzpatrick:2016mtp}.
The Virasoro block with the intermediate operator $\mathcal{O}_p$ was similarly obtained up to $1/c$ order as
\begin{align}
\begin{aligned}
z^{2 h_j} \mathcal{V}_p (z)=& z^{h_p} {}_2 F_1 (h_p , h_p ;2 h_p ; z) \\
&+ \frac{1}{c} \left[ h_j h_q f_a (h_p, z) + (h_j + h_q ) f_b (h_p, z) + f_c (h_p, z ) \right] + \mathcal{O}(c^{-2})  \, .
\end{aligned}
\label{Vp}
\end{align}
All order expressions in $z$  for $f_a$ and $f_b$ and the first few orders in $z$ for $f_c$ were obtained in \cite{Fitzpatrick:2016mtp}. In subsection \ref{GenericCB}, we examine the general block, applying our method, and, in particular, we obtain all order expressions in $z$ even for $f_c$.

For correlators in the W$_N$ minimal model with general $N$, we may compute correlators by applying the Coulomb gas method as explained in appendix \ref{toda}, see, e.g., \cite{Fateev:2007ab,Fateev:2008bm,Papadodimas:2011pf,Chang:2011vka,Hijano:2013fja}.
A conformal block is given by a choice of integration contour over the positions of screening operators inserted. Since it is non-trivial to identify proper integration contours,%
\footnote{We are particularly interested in the identity W$_N$ block, which appears in the $s$-channel decomposition of the four point function \eqref{4pt0}. However, we cannot find out any references giving the $s$-channel expressions. The expressions in the $t$-channel or the $u$-channel were obtained in \cite{Fateev:2007ab,Fateev:2008bm} when $\Lambda$ or $\Xi$ correspond to a degenerate representation, see also \cite{Chang:2011vka}.}
we consider the simple correlator 
\begin{align}
\langle
\bar{\cO}_{(n \bm{\omega}_m ; 0)}(\infty) \cO_{(n \bm{\omega}_m ; 0)}(1)  \bar{\cO}_{(\text{f}; 0)}(z) \cO_{(\text{f}; 0)}(0)\rangle \, .
\label{4pt1}
\end{align}
In this case, the Coulomb integrals reduce to a hypergeometric function ${}_2 F_1 (a,b;c;z)$, and we can apply the analysis for the Virasoro minimal model, see, e.g., \cite{DiFrancesco:1997nk}.
The final result obtained in appendix \ref{toda} is
\begin{align}
& |z|^{- 4 h (\text{f};0)} |1-z|^{2 nm\frac{k+N+1}{N(k+N)}} \biggl[ \left| {}_2F_1 \left(n + \frac{n}{k+N},-\frac{m}{k+N};-\frac{N}{k+N};z \right) \right|^2 
\label{coulomb}
\\
&   + \cN \left|  z^{\frac{k+2N}{k+N}} (1 -z)^{\frac{m - N - n (k + N +1)}{k+N}} {}_2 F_1 \left (1+\frac{m}{k+N},-n+1-\frac{n}{k+N};2 + \frac{N}{k+N};z \right) \right|^2  \biggr]  \nonumber
\end{align}
with 
\begin{align}
\begin{aligned}
\cN \equiv -\frac{\Gamma(1+\frac{m}{k+N}) \Gamma(-n+1-\frac{n}{k+N}) \Gamma(1+\frac{N-m}{k+N}) \Gamma(n+1+\frac{N+n}{k+N}) \Gamma(-\frac{N}{k+N})^2}{\Gamma(n+\frac{n}{k+N}) \Gamma(-\frac{m}{k+N}) \Gamma(-n-\frac{N+n}{k+N}) \Gamma(-\frac{N-m}{k+N}) \Gamma(2+\frac{N}{k+N})^2} \;.
\end{aligned}
\label{CoulombN}
\end{align}
The first and second terms correspond to the W$_N$ blocks of identity and adjoint operators, respectively. In subsection \ref{VacuumCB}, we examine the identity W$_3$ block up to the $1/c^2$ order. In subsection \ref{largec}, we give some comments on general W$_3$ blocks, and in appendix \ref{simple}, we analyze the adjoint W$_3$ block with $n=m=1$ up to the $1/c$ order. 

The identity W$_N$ block is a useful quantity even in the CFT perspective. 
For instance, we proposed a way to obtain the $1/c$ correction of spin $s$ charge for the primary operator in the W$_N$ minimal model \eqref{coset} by utilizing the Virasoro block decomposition of the identity W$_N$ block in \cite{Hikida:2017byl,Hikida:2017ehf}. Since the W$_N$ minimal model is a solvable model,
we know how to obtain the higher spin charges with finite $N,k$ as in \cite{Bais:1987zk,Ahn:2011by,Ahn:2013sua}. However, in practice, it is difficult to obtain their explicit expressions for larger $s$, and our method provides a simple way to obtain them including $1/c$ corrections.
In \cite{Hikida:2017byl,Hikida:2017ehf}, we have considered higher spin charges for operators only in the fundamental representation. In appendix \ref{toda}, we extend the analysis by examining higher spin charges for the operator with $\Lambda = n \bm{\omega}_m$ including $1/c$ corrections.%
\footnote{See \cite{Ahn:2017noo} for a recent work on higher spin charges for general operators in a supersymmetric model.}
Among others, we can reproduce the $1/c$ correction of spin 3 charge in \eqref{wexpN}.

We also apply our approach to heavy-light correlators.
At the leading order in $1/c$, general CFT correlators of this type were reproduced from the analysis of the open Wilson line in a conical space in \cite{Hijano:2013fja,Hegde:2015dqh}.
For the analysis of $1/c$ corrections, we may work on a simple form of the correlator as
\begin{align}
\langle
\bar{\cO}_{(0; n \bm{\omega}_m )}(\infty) \cO_{(\text{f}; 0)}(1) \bar{\cO}_{(\text{f} ; 0)}(z)  \cO_{(0; n \bm{\omega}_m)}(0) \rangle \, .
\label{HLN}
\end{align}
In this case, there is only one type of exchanged operator, and the Coulomb integrals can be carried out as (see, e.g.,  \cite{Hijano:2013fja})
\begin{align}
|1 - z|^{- 4 h(\text{f};0)}| z |^{ - \frac{2  n m }{N}}
 \left|{}_2 F_1 \left(-n, - \frac{m}{N+k};- \frac{N}{N+k}; 1 - z \right)  \right|^2 \, .
\label{HLN2}
\end{align}
We reproduce the CFT result for $N=2$ by applying our method for open Wilson lines and the analysis for conical spaces in  \cite{Raeymaekers:2014kea}.
The bulk approach should  be applicable also for general $N$, but we leave detailed analysis to future work.

\section{Method for bulk computation}
\label{BulkMethod}

In this section, we explain our method to compute correlators or conformal blocks from the networks of open Wilson lines. 
In the next subsection, we first introduce $\text{sl}(N) \oplus \text{sl}(N)$ Chern-Simons gauge theory and the networks of Wilson lines corresponding to conformal blocks. We then explain  our prescription developed in \cite{Hikida:2017ehf} to remove divergences associated with loop diagrams. 
In subsection \ref{23ptN2}, we review our previous work in \cite{Hikida:2017ehf} on the two point function of a scalar operator and the three point function with two scalars and a spin 2 current for $N=2$.
The analysis can be extended straightforwardly for $N=3$ if the scalar operator belongs to the (anti-)fundamental representation of su(3).
In subsection \ref{WilsonNew}, we extend our method by analyzing correlators involving general scalar operators.

\subsection{Chern-Simons theory and open Wilson lines}
\label{Wilson}

The bulk description of the W$_N$ minimal model with large $c$ was proposed to be given by $\text{sl}(N) \oplus \text{sl}(N)$ Chern-Simons gauge theory \cite{Castro:2011iw},
the action of which is
\begin{align}
S = S_\text{CS} [A] - S_\text{CS} [\tilde A] \, , \quad
S_\text{CS} [A] = \frac{\hat k}{4 \pi} \int \text{tr} \left( A \wedge d A + \frac23 A \wedge A \wedge A \right) \, . \label{CSaction}
\end{align}
Here, $\hat k$ is the level of the Chern-Simons theory, and $A$ takes values in sl$(N)$ Lie algebra.%
\footnote{We neglect $\tilde A$ throughout this paper, but $\tilde A$ can be analyzed similarly.}
As a higher spin gravity, it is important to decompose the generators of sl$(N)$ Lie algebra in terms of  embedded sl$(2)$. We consider the principal embedding as
\begin{align}
\text{sl}(N) = \text{sl}(2) \oplus \left( \bigoplus_{s=3}^N g^{(s)}\right)  \, .
\end{align}
Here, $g^{(s)}$ denotes the $(2s - 1)$-dimensional representation of sl$(2)$, which is generated by
$V_{n}^{s}$ $(n=-s+1,-s+2 , \ldots , s-1)$.
A solution to the equations of motion can be written in a form as
\begin{align}
A = e^{- \rho V_0^2} a(z) e^{\rho V_0^2} dz + V_0^2 d \rho
\end{align}
by fixing a gauge. Here, $\rho$ denotes the radial coordinate, and the boundary is located at 
$\rho = \infty$.
Moreover, $z , \bar z$ are coordinates parallel to the boundary.
With the coordinates, the AdS metric is given by $ds^2 = d \rho^2 + e^{2 \rho} dz d\bar z$, and the corresponding configuration of the gauge field is with $a(z) = V_1^{2}$.
As for the application to AdS gravity, we assign the asymptotic AdS condition, which restricts the from of the gauge field as
\cite{Henneaux:2010xg,Campoleoni:2010zq,Gaberdiel:2011wb,Campoleoni:2011hg}
\begin{align}
a (z) = V_1^{2} - \frac{1}{\hat k} \sum_{s = 2}^N \frac{1}{N_s} J^{(s)} (z) V^{s}_{-s +1}
\label{az}
\end{align}
with $N_s$ in \eqref{BsNs}.
The asymptotic symmetry near the AdS boundary is identified as the W$_N$ algebra generated by $J^{(s)}$.
The central charge of the algebra is
\begin{align}
c = 6 \hat k
\label{BH}
\end{align}
at the classical level, which is same as for the pure gravity in \cite{Brown:1986nw}.
With the coefficients of $J^{(s)}$ in \eqref{az}, the normalization of currents is the same as that in  \eqref{2ptcurrent} at the leading order in $1/c$.

We would like to compute the correlators or conformal blocks of the W$_N$ minimal model in terms of Wilson line operator
\begin{align}
W (z_f ; z_i) = P \exp\left ( \int_{z_i}^{z_f} d z \,  a(z)  \right)
\label{WL}
\end{align}
in the sl$(N)$ Chern-Simons theory.
Here we remove the $\rho$-dependence by a gauge transformation as $A = a(z)$.
At the leading order in $1/c$, it has been argued that $n$-point conformal blocks can be evaluated as \cite{Bhatta:2016hpz,Besken:2016ooo,Besken:2017fsj}
\begin{align}
G_n(\Lambda_j|z_j) \equiv  \langle S |  \prod_{j = 1}^n  W_{\Lambda_j} (z_0 ; z_j)   |\text{hw}\rangle _{j} \, ,
\label{WLN}
\end{align}
where $  | \text{hw} \rangle _{j} $ denotes the highest weight state in the representation $\Lambda_j$ of sl$(N)$.
Moreover, $\langle S |$ belongs to a singlet representation in $\otimes_{j=1}^n \Lambda_j$.
There are several ways to construct a singlet state, and a choice of singlet leads to one of conformal blocks, see \cite{Besken:2016ooo} for more explanations and explicit examples.
The $n$ open Wilson lines are connected at a point $z_0$, but the expression should be independent of $z_0$. For $n=2$, we may set $z_0 =z_2$,%
\footnote{If we instead set $z_0 =z_1$, then we obtain $
	G_2 (\Lambda,\bar \Lambda | z_1 , z_2) =  \langle \text{lw} | W_{\bar{ \Lambda}} (z_1 ; z_2) | \text{hw} \rangle $ with $ | \text{hw} \rangle $ and $\langle \text{lw} |$ in the representations $\bar \Lambda$  and $\Lambda$, respectively.
}
then, the expression of \eqref{WLN} becomes
\begin{align}
G_2 ( \Lambda, \bar \Lambda | z_1 , z_2) =  \langle \text{lw} | W_{ \Lambda} (z_2 ; z_1) | \text{hw} \rangle \, .
\label{G2}
\end{align}
Here $ \langle \text{lw} | $ represents the lowest weight state in the representation $\bar \Lambda$, which is the conjugate of $ \Lambda$.

For evaluating quantum effects in the Wilson line network \eqref{WLN}, we should integrate over gauge fields in the path integral with the insertions of Wilson line operators. Following \cite{Fitzpatrick:2016mtp,Besken:2017fsj,Hikida:2017ehf}, we evaluate the expectation values by using the correlators of $J^{(s)}(z)$ in \eqref{WL} with \eqref{az} in terms of the theory with W$_N$ symmetry as in \eqref{2ptcurrent}. Here the normalization \eqref{BsNs} is set in terms of the central charge $c$, so it is regarded as the full propagator after including loop corrections, see \cite{Besken:2017fsj} for the arguments. We also need to integrate over $z$ along the Wilson line in \eqref{WL}. However, the integration would diverge when at least two of $J^{(s)}(z)$ collide at the same point. As in \cite{Hikida:2017ehf}, we introduce a regular $\epsilon$ and replace \eqref{2ptcurrent} by
\begin{align}
\langle J^{(s)} (z_1) J^{(s)} (z_2) \rangle = \frac{B_s}{z_{12}^{2 s  - 2 \epsilon}} \, .
\label{2ptcurrentr}
\end{align}
A merit of this choice is that the regulator does not break the scale invariance. The integration over $z_i$ would diverge when $\epsilon \to 0$, and we remove the divergence by renormalizing the overall normalization of Wilson lines and  parameters corresponding to interactions as in \cite{Hikida:2017ehf}. We denote the parameters as 
\begin{align}
c_s = 1 + \frac{1}{c} c_s^{(1)} + \mathcal{O} (c^{-2}) \, , 
\end{align} 
and insert them into the Wilson line operator as
\begin{align}
W (z_f;z_i)
= P \exp \left[ \int_{z_i}^{z_f} dz \left(  V_1^{2} - \frac{6}{c} \sum_{s=2}^N \frac{c_s}{N_s} J^{(s)} (z) V_{-s+1}^{s} \right) \right] \, .
\label{ren}
\end{align}
We can remove some divergences at the $1/c$ order by using $c_s^{(1)}$ and similarly for higher orders. There are still ambiguities for the terms of order $\epsilon^0$, and we offer to fix them such that the boundary correlators are consistent with the W$_N$ symmetry.

\subsection{Two and three point functions for $N=2$}
\label{23ptN2}

In order to illustrate our method and fix notations, we review our analysis in \cite{Hikida:2017ehf} for the two and three point functions in \eqref{23pt}.
For $N=2$, it is convenient to work with the $x$-basis, which can be applied to the  general $N$ case but with operators in the (anti-) fundamental representation. As in \cite{Verlinde:1989ua,Fitzpatrick:2016mtp,Hikida:2017ehf}, we consider the $x$-basis expression of \eqref{G2} as
\begin{align}
 \langle \text{lw}| W_{h_0} (z_f ; z_i)  | \text{hw} \rangle
 =
\int d x\langle \text{lw} | x \rangle  W_{h_0} (z_f ; z_i) \langle x | \text{hw} \rangle \, .
\label{G2x}
\end{align}
For brevity, we call the left hand side $W_{h_0} (z_f ; z_i) $.
With the basis, we can use
\begin{align}
\langle \text{lw} | x  \rangle = \delta (x) \, , \quad
\langle x | \text{hw} \rangle = \frac{1}{x^{2 h_0}} \, ,
\label{statesN2}
\end{align}
where $h_0 = -j$ for $N=2$ and $h_0 = (1-N)/2$ for general $N$. The generators are \cite{Bergshoeff:1991dz}
\begin{align}
V_n^s = \sum_{i=0}^{s-1}\begin{pmatrix} s -1 \\ i \end{pmatrix} \frac{(n - s + 1)_{s - 1 - i} (- 2 h_0 - s + 2)_{s - 1 - i}}{(s+i)_{s-1-i}} x^{- n + i} \partial^i_x \, ,
\end{align}
and, in particular, the sl$(2)$ subalgebra is generated by
\begin{align}
V_1^2 = \partial_x \, , \quad
V_0^2 = x \partial_x + h_0 \, , \quad
V_{-1}^2 = x^2 \partial_x + 2 h_0 x \, .
\label{sl2gen}
\end{align}
As in \cite{Fitzpatrick:2016mtp,Besken:2017fsj,Hikida:2017ehf}, we express \eqref{G2x} as
\begin{align}
\begin{aligned}
W_{h_0} (z_f ; z_i) = &\sum_{n=0}^\infty \left( - \frac{6}{c}\right)^n \int_{z_i}^{z_f} dz_n \cdots
\int_{z_i}^{z_2} dz_1
\\
& \times \sum_{s_j = 2}^N \left [ \prod_{j=1}^n \frac{c_{s_j}}{N_{s_j}} J^{(s_j)} (z_j)\right ] f_{h_0}^{(s_n , \ldots , s_1)} (z_f;z_i;z_n , \ldots , z_1)
\end{aligned}
\label{Wh}
\end{align}
with $(X=x)$
\begin{align}
&f_{h_0}^{(s_n , \ldots , s_1)} (z_f;z_i;z_n , \ldots , z_1)\label{fh}
 \\
&\quad = \left. \left[ e^{(z_f - z_n) V^2_1 } V^{s_n}_{-s_n +1} e^{ (z_n - z_f) V^2_1} \right] \cdots \left[ e^{(z_f - z_1) V^2_1 } V^{s_1}_{-s_1 +1} e^{ (z_1 - z_f) V^2_1} \right] e^{(z_f - z_i) V^2_1 }
 \langle X | \text{hw} \rangle   \right|_{X=0} \nonumber
\end{align}
 in $1/c$ expansion.

Let us first evaluate the two point function of operator $\mathcal{O}_j$ for $N=2$  as
\begin{align}
\langle \mathcal{O}_j (z) \mathcal{O}_j (0) \rangle = \langle W_{h_0} (z;0) \rangle
\end{align}
with $h_0 = - j$ by applying our Wilson line method.
At the leading order in $1/c$, the two point function becomes
\begin{align}
\left. \langle W_{h_0} (z;0) \rangle   \right|_{\mathcal{O}(c^0)} =  \frac{1}{z^{ 2 h_0}} 
\label{2ptN2tree}
\end{align}
as expected.
The contribution at the next order is
\begin{align}
\begin{aligned}
\left. \langle W_{h_0} (z;0) \rangle   \right|_{\mathcal{O}(c^{-1})} & = \left( \frac{6}{c}\right)^2 \int_0^z dz_2 \int_0^{z_2} d z_1  f_{h_0}^{(2,2)} (z;0;z_2,z_1) \langle J^{(2)} (z_2) J^{(2)}(z_1) \rangle \\
 & = \frac{1}{c z^{2 h_0}} \left[ \frac{6 (h_0 - 1)h_0}{\epsilon} + 12 h_0 (h_0 - 1) \log(z) + 2 h_0 (5 h_0 - 2) \right]  \, ,
\end{aligned}
\end{align}
where we have used \eqref{2ptcurrentr}.
We can remove the term proportional to $1/\epsilon$ by changing the overall normalization of Wilson line operator. The shift of conformal weight can be read off from the term proportional to $\log (z)$ as in \eqref{cwshift}, and the $1/c$ order term in \eqref{hN2} is successfully reproduced.

We also examine a three point function using Wilson line operator as
\begin{align}
\langle \mathcal{O}_j (z)\mathcal{O}_j (0)J^{(2)}(y)\rangle = \langle W_{h_0} (z;0) J^{(2)}(y) \rangle
\end{align}
with the extra insertion of the spin 2 current. At the leading order in $1/c$, we have
\begin{align}
\begin{aligned}
&\lim_{y \to - \infty} |y|^4 \left. \langle W_{h_0} (z;0) J(y) \rangle\right|_{\mathcal{O}(c^0)} \\
&= \lim_{y \to - \infty} |y|^4 \frac{6}{c} \int_0^z dz_1 f^{(2)}_{h_0} (z;0;z_1)\langle J^{(2)} (z_1) J^{(2)} (y ) \rangle
= h_0 z^2 \left. \langle W_{h_0} (z;0) \rangle \right|_{\mathcal{O}(c^0)}
\end{aligned}
\end{align}
as in \eqref{3pt}. At the next leading order, there are two types of contributions as
\begin{align}
&  \left(\frac{6}{c} \right)^2  \int_0^z dz_2 \int_0^{z_2} dz_1 f^{(2,2)}_{h_0} (z;0;z_2 , z_1)\langle J^{(2)} (z_2)  J^{(2)} (z_1) J^{(2)} (y ) \rangle \, , \\
 &  \left(\frac{6}{c} \right)^3  \int_0^z dz_3 \int_0^{z_3} dz_2 \int_0^{z_2} dz_1 f^{(2,2,2)}_{h_0} (z;0;z_3, z_2 , z_1)\langle J^{(2)} (z_3) J^{(2)} (z_2)  J^{(2)} (z_1) J^{(2)} (y ) \rangle \, . \nonumber
\end{align}
We use the correlators
\begin{align}
\langle J^{(2)} (z_3) J^{(2)} (z_2) J^{(2)} (z_1) \rangle
 = \frac{c}{(z_{21}  z_{32}  z_{31})^{2 - \epsilon}}
 \label{3ptspin2}
\end{align}
for the three point function and
\begin{align}
\langle J^{(s)} (z_4) J^{(s)} (z_3) J^{(s)} (z_2) J^{(s)} (z_1) \rangle
= \frac{B_s^2}{(z_{43}  z_{21})^{2 s - 2 \epsilon}} + \frac{B_s^2}{(z_{42}  z_{31})^{2 s - 2 \epsilon}} + \frac{B_s^2}{(z_{41}  z_{32})^{2 s - 2 \epsilon}}
\label{4ptspins}
\end{align}
with $s=2$ up to order $c^2$ for the four point function.
The sum of the integrals leads to
\begin{align}
\lim_{y \to - \infty} |y|^{4 - 2 \epsilon}  \langle W_{h_0} (z;0) J^{(2)}(y) \rangle 
=\left[ h_0- \frac{6 h_0}{c} \left( \frac{1}{\epsilon} + h_0 - \frac12 \right) \right]   z^2  \langle W_{h_0} (z;0) \rangle + \cdots \, .
\end{align}
We remove the divergence by renormalizing the coupling $c_2$ in \eqref{ren} as (see (4.21) of \cite{Hikida:2017ehf})
\begin{align}
c_2 = 1 + \frac{1}{c} \left( \frac{6}{\epsilon} + 3 \right) + \mathcal{O}(c^{-2}) \, .
\label{c2reg}
\end{align}
Here the constant term is chosen such that the three point function becomes
\begin{align}
\lim_{y \to - \infty} |y|^4 \langle W_{h_0} (z;0) J^{(2)}(y) \rangle
= h_jz^2 \langle W_{h_0} (z;0) \rangle
\end{align}
with $h_j$ in \eqref{hN2} up to order $1/c$. In other words, we chose the renormalization scheme to be consistent with the conformal Ward identity in the boundary theory.
Here, we remark that the renormalized coupling $c_2$ in \eqref{c2reg} is independent of $h_0$; thus,  the same Wilson line operator \eqref{ren} can be used to evaluate the network in \eqref{WLN}.

In \cite{Hikida:2017ehf}, we also examined the $1/c^2$ order corrections of the two point function, and we reproduced the shift of conformal weight at the $1/c^2$ order as in \eqref{hN2}. This is possible only after properly regularizing divergences and this fact gives a support for our renormalization prescription, see \cite{Hikida:2017ehf} for more details. The analysis here can be directly applied to the case with $N=3$ but involving scalar operators only in the (anti-)fundamental representation. Similarly for $N=2$, we showed in \cite{Hikida:2017ehf} that the divergences at the $1/c$ order in the two and three point functions of the form \eqref{23pt} can be removed by renormalizing the overall factor of Wilson line operator and couplings $c_2$ and $c_3$ introduced as in \eqref{ren}. With the renormalization at the $1/c$ order, we reproduced the shift of conformal weight of scalar operator up to the $1/c^2$ order.

\subsection{Two and three point functions for $N=3$}
\label{WilsonNew}

In \cite{Hikida:2017ehf}, we restrict ourselves to operators only in the (anti-)fundamental representation. However, the fusion of this type of operators would yield another operator belonging to a representation other than the (anti-)fundamental one. Therefore, it is necessary to extend the previous analysis and deal with correlators involving general operators.
It is known that sl$(N)$ generators acting on general states can be represented by using parameters $x_a$ $(a = 1,2, \ldots, (N-1)N/2)$ with the number of the positive root.
For $N=3$, we introduce three parameters $X=(x,y,w)$, which makes the expression \eqref{Wh} with \eqref{fh} applicable to the general situation. The explicit expressions of states and generators are summarized in appendix \ref{sl3}.

Just like the $N=2$ case, we start from computing the two point function of operator $\mathcal{O}_{(\lambda_1 , \lambda_2 ; 0)}$ using the Wilson line as
\begin{align}
	\langle \bar{\mathcal{O}}_{(\lambda_1 , \lambda_2 ; 0)} (z)  \mathcal{O}_{(\lambda_1 , \lambda_2 ; 0)} (0) \rangle = \langle  W_{(\lambda_1 , \lambda_2)} (z ; 0) \rangle \, .
\end{align}
The expression of the Wilson line in $1/c$ expansions in \eqref{Wh} with \eqref{fh} leads to
\begin{align}
 \left.	\langle  W_{(\lambda_1 , \lambda_2)} (z) \rangle \right|_{\mathcal{O}(c^0)}
  =  z^{2 \lambda_1 + 2 \lambda_2} 
  \label{2ptN3tree}
\end{align}
with the normalization of \eqref{Xhw}.
This is consistent with the leading order expression of conformal weight in \eqref{hexp}.
At the next leading order in $1/c$, we have the contributions as $(s=2,3)$
\begin{align}
\left( \frac{6}{c}\right)^2  \int_0^z dz_2 \int_0^{z_2} dz_1 f^{(s,s)}_{(\lambda_1 ,\lambda_2)} (z;0;z_2 ,z_1)
	\langle J^{(s)} (z_2) J^{(s)} (z_1) \rangle \, ,
\end{align}
where $f_{(\lambda_1, \lambda_2)}^{(s,s)} (z;0;z_2,z_1)$ are given by \eqref{fh} with \eqref{xshift}, \eqref{gshift}, and \eqref{gshift2}. Moreover, we use the two point functions of currents  in \eqref{2ptcurrentr} with
\begin{align}
B_2 = \frac{c}{2} \, , \quad 
B_3 = - \frac{5 c}{6} \, ,
\label{B2B3}
\end{align}
see \eqref{B2B30} with $N=3$.
The sum of the both contributions is
\begin{align}
\begin{aligned}
	\langle W_{(\lambda_1 , \lambda_2)} (z) \rangle =&  z^{2 \lambda_1 + 2 \lambda_2}  + \frac{1}{c} z ^ { 2 \lambda_1 + 2 \lambda_2} \Biggl[
	\frac{8 \left(\lambda_1^2+\lambda_1 (\lambda_2+3)  +\lambda_2 (\lambda_2+3)\right) }{\epsilon}  \\ &+\frac{1}{15} \left(240 \left(\lambda_1^2+\lambda_1 (\lambda_2+3)+\lambda_2 (\lambda_2+3)\right) \log (z) \right. \\ & \left. +197 \lambda_1^2+\lambda_1 (206 \lambda_2+213)+\lambda_2 (197 \lambda_2+213)\right) \Biggr] 
\end{aligned}
\end{align}
up to the orders of $\epsilon^0$ and $1/c$. The $z$-independent overall factor can be set as $1$ by changing the normalization of the Wilson line operator.
From the coefficient of  $\log(z)$, the shift of conformal dimension can be read off as
\begin{align}
	- \frac12  \cdot  \frac{1}{c} \cdot 16 \left(\lambda_1^2+\lambda_1 (\lambda_2+3)+\lambda_2 (\lambda_2+3)\right)  \, ,
\end{align}
which reproduces the $1/c$ order term in  \eqref{hexp}.

Next we move to the evaluation of Wilson line operator with the insertion of a higher spin current $J^{(s)}$ with $s=2$ or $s=3$.
We would like to show that the three point functions can be reproduced as
\begin{align}
\langle \bar{\mathcal{O}}_{(\lambda_1 , \lambda_2 ; 0)} (z)  \mathcal{O}_{(\lambda_1 , \lambda_2 ; 0)} (0) J^{(s)} (y) \rangle =   \langle  W_{(\lambda_1 , \lambda_2)} (z ;0) J^{(s)} (y) \rangle \, .
\end{align}
Let us define
\begin{align}
G_3^{(s)} (z) = \lim_{y \to - \infty} |y|^{2 s} \langle W_{(\lambda_1,\lambda_2)} (z) J^{(s)} (y)\rangle
\label{G3s}
\end{align}
with $s=2,3$. We first examine the leading order in $1/c$.
From the expression of the open Wilson line in \eqref{Wh} with \eqref{fh}, we find
\begin{align}
\begin{aligned}
& \left. G_3^{(2)} (z)  \right|_{\mathcal{O}(c^0)}
= \frac{6}{c} B_2 \int_0^z dz_1 f^{(2)}_1 (z_1)
=  - (\lambda_1 + \lambda_2) z^{2 + 2 \lambda_1  + 2 \lambda_2} \, , \\
& \left. G_3^{(3)} (z)  \right|_{\mathcal{O}(c^0)}
= -  \frac{6}{c} B_3\int_0^z dz_1 f^{(3)}_1 (z_1)
= \frac13  (\lambda_1 - \lambda_2) z^{3 + 2 \lambda_1  + 2 \lambda_2}
\end{aligned}
\end{align}
using \eqref{xshift}, \eqref{gshift}, and \eqref{gshift2}.
There are no divergences from these tree level computations, which reproduce \eqref{3pt} with \eqref{hexp} and \eqref{wexp} at this order.

Divergences arise from the next order in $1/c$.
For $G_3^{(2)} (z)$ in \eqref{G3s} at this order, a type of contributions comes from the integrals as%
\footnote{We take the limit $
	G_3^{(s)} (z) |_{\mathcal{O}(c^{-1})} = \lim_{y \to - \infty} |y|^{2 s - 2 \epsilon} \langle W_{(\lambda_1,\lambda_2)} (z) J^{(s)} (y)\rangle |_{\mathcal{O}(c^{-1})} $ for these computations.}
\begin{align}
\begin{aligned}
&\left( \frac{6}{c}\right)^2 c \int_0^z dz_2 \int_0^{z_2}  dz_1 f_{(\lambda_1,\lambda_2)}^{(2,2)} (z;0 ;z_2,z_1) \frac{1}{z_{21}^{2 - \epsilon}} \, , \\
&- \left( \frac{6}{c}\right)^2 \frac{5 c}{2} \int_0^z dz_2 \int_0^{z_2}  dz_1 f_{(\lambda_1,\lambda_2)}^{(3,3)} (z;0 ; z_2,z_1) \frac{1}{z_{21}^{4 -  \epsilon}} \, .
\end{aligned}
\end{align}
Here, we have used the three point functions \eqref{3ptspin2} and
\begin{align}
&\langle J^{(3)} (z_3) J^{(3)} (z_2) J^{(2)} (z_1) \rangle = \frac{- 5c/2}{z_{32}^{4 - \epsilon} z_{21}^{2 - \epsilon}z_{31}^{2 - \epsilon}} \, , \label{3ptcurrent}
\end{align}
where the powers are obtained by shifting the conformal weights of $J^{(2)}$ and $J^{(3)}$ as $2 \to 2 - \epsilon$ and $3 \to 3 - \epsilon$, respectively.
Another type is
\begin{align}
\begin{aligned}
&\left( \frac{6}{c}\right)^3 B_s B_2  \int_0^z dz_3 \int_0^{z_3} dz_2 \int_0^{z_2} dz_1 f_{(\lambda_1,\lambda_2)}^{(s,s,2)} (z;0;z_3,z_2,z_1)\frac{1}{z_{32}^{2s - 2 \epsilon}} \, ,  \\
&\left( \frac{6}{c}\right)^3 B_s B_2  \int_0^z dz_3 \int_0^{z_3} dz_2 \int_0^{z_2} dz_1 f_{(\lambda_1,\lambda_2)}^{(2,s,s)} (z;0;z_3,z_2,z_1)\frac{1}{z_{21}^{2s - 2 \epsilon}} \, , \\
&\left( \frac{6}{c}\right)^3 B_s B_2  \int_0^z dz_3 \int_0^{z_3} dz_2 \int_0^{z_2} dz_1 f_{(\lambda_1,\lambda_2)}^{(s,2,s)} (z;0;z_3,z_2,z_1)\frac{1}{z_{31}^{2s - 2 \epsilon}}  \end{aligned}
\end{align}
with $s=2,3$.  Here the large $c$ factorization of four point functions is used as \eqref{4ptspins} and
\begin{align}
\langle J^{(3)} (z_4) J^{(3)} (z_3) J^{(2)} (z_2) J^{(2)} (z_1) \rangle
= \frac{B_2 B_3}{z_{43} ^{6 - 2\epsilon}z_{21}^{4 -2 \epsilon}} + \mathcal{O}(c) \, . \label{4ptcurrent}
\end{align}
Totally, we have
\begin{align}
\begin{aligned}
&z^{-2} G_3^{(2)} (z)
=\Biggl[ - (\lambda_1 + \lambda_2)
\\ & \quad + \frac{1}{c}\left( \frac{36 (\lambda_1+\lambda_2) }{\epsilon}-\left(8 \lambda_1^2+\lambda_1 (8 \lambda_2+11)+\lambda_2 (8 \lambda_2+11)\right)  \right) \Biggr]  \langle W_{(\lambda_1 , \lambda_2)} (z;0)\rangle
\end{aligned}
\end{align}
up to orders $\epsilon^0$ and $1/c$.
There is  a divergent term proportional to $1/\epsilon$, which is removed by shifting the coupling $c_2$ introduced in \eqref{ren}  from its classical value $c_2 = 1$.
We choose the $\epsilon$-independent part of $c_2$ as
\begin{align}
c_2 = 1 + \frac{1}{c} \left( \frac{36}{\epsilon} + 13\right) + \mathcal{O} (c^{-2}) \, ,
\label{c2shift}
\end{align}
then we  reproduce the CFT result in \eqref{3pt} with \eqref{hexp} as
\begin{align}
z^{-2} G_3^{(2)} (z) & = h(\lambda_1 , \lambda_2 ;0)\langle W_{(\lambda_1 , \lambda_2)} (z)\rangle
\end{align}
up to the order $1/c$.
Note that the shift of coupling  $c_2$ in \eqref{c2shift} is the same as (5.17) of \cite{Hikida:2017ehf}, and it is independent of the representation $(\lambda_1,\lambda_2)$.

We can similarly analyze $G_3^{(3)}(z)$ in \eqref{G3s}.
At the $1/c$ order, contributions are
\begin{align}
& \left( \frac{6}{c}\right)^2  \frac{5 c}{2} \int_0^z dz_2 \int_0^{z_2}  dz_1 \left[ f_{(\lambda_1,\lambda_2)}^{(2,3)} (z;0,z_2,z_1) + f_{(\lambda_1,\lambda_2)}^{(3,2)} (z;0,z_2,z_1) \right] \frac{1}{z_{21}^{2 - \epsilon}} \, ,
\end{align}
and
\begin{align}
\begin{aligned}
&-\left( \frac{6}{c}\right)^3 B_s B_3\int_0^z dz_3 \int_0^{z_3} dz_2 \int_0^{z_2} dz_1 f_{(\lambda_1,\lambda_2)}^{(s,s,3)} (z;0,z_3,z_2,z_1)\frac{1}{z_{32}^{2s - 2 \epsilon}} \, ,  \\
&-\left( \frac{6}{c}\right)^3 B_s B_3 \int_0^z dz_3 \int_0^{z_3} dz_2 \int_0^{z_2} dz_1 f_{(\lambda_1,\lambda_2)}^{(3,s,s)} (z;0,z_3,z_2,z_1)\frac{1}{z_{21}^{2s - 2 \epsilon}} \, , \\
&-\left( \frac{6}{c}\right)^3 B_s B_3 \int_0^z dz_3 \int_0^{z_3} dz_2 \int_0^{z_2} dz_1 f_{(\lambda_1,\lambda_2)}^{(s,3,s)} (z;0,z_3,z_2,z_1)\frac{1}{z_{31}^{2s - 2 \epsilon}}  \end{aligned}
\end{align}
with $s=2,3$. Here we have used \eqref{3ptcurrent}, \eqref{4ptspins}, and \eqref{4ptcurrent}.
 The sum of them leads to
\begin{align}
\begin{aligned}
&z^{-3} G_3^{(3)} (z)  =\Biggl[  \frac{\lambda_1 - \lambda_2}{3}  \\ & \quad
+ \frac{1}{c}\left( -\frac{12 (\lambda_1-\lambda_2) }{\epsilon} + \frac{1}{5} (\lambda_1-\lambda_2) (40 \lambda_1+40 \lambda_2+33)  \right) \Biggr]  \langle W_{(\lambda_1 , \lambda_2)} (z)\rangle  \end{aligned}
\end{align}
up to orders $\epsilon^0$ and $1/c$. Changing  the coupling  $c_3$ in \eqref{ren} as (see (5.30) of \cite{Hikida:2017ehf})
\begin{align}
c_3 =  1 + \frac{1}{c} \left( \frac{36}{\epsilon} + 1\right) + \mathcal{O} (c^{-2}) \, ,
\end{align}
 the above expression becomes
\begin{align}
z^{-3} G_3^{(3)} (z) & = w(\lambda_1 , \lambda_2 ;0)\langle W_{(\lambda_1 , \lambda_2)} (z;0)\rangle
\end{align}
up to the $1/c$ order as in \eqref{3pt} with \eqref{wexp}. Here, we  remark that the shift of $c_3$ is independent of $\lambda_1,\lambda_2$ as for $c_2$.

Adopting the regularization, we should be able to reproduce the shift of conformal weight up to $1/c^2$ order just like the case with $(\lambda_1,\lambda_2) = (1,0)$ in \cite{Hikida:2017ehf}. Here we do not repeat the computation since it is not necessary for the following analysis.

\section{Conformal blocks from open Wilson lines}
\label{CBWilson}

In this section, we evaluate four point conformal blocks of the form \eqref{4pt0}  by including quantum effects on the networks of open Wilson lines. 
For $N=2$, similar computations were done in \cite{Fitzpatrick:2016mtp}, but their formulation is  different from ours. 
In the next subsection we start from the identity blocks, which are computable by the products of two open Wilson lines. Since computations turn out to be quite similar to those in \cite{Fitzpatrick:2016mtp} for $N=2$, we focus on the $N=3$ case by applying the method developed in the previous section. In subsection \ref{GenericCB}, we examine general conformal blocks, where we have to really deal with the networks of open Wilson lines. 
We concentrate on the simpler case with $N=2$ and briefly discuss the $N=3$ extensions, see also appendix \ref{simple}.

\subsection{Identity W$_3$ blocks}
\label{VacuumCB}

In order to express a conformal block for \eqref{4pt0} in terms of the Wilson line network as in \eqref{WLN}, we have to choose a proper singlet $ S $ from the product of representations 
$\bar{\Lambda} \otimes \Lambda \otimes \bar{\Xi} \otimes \Xi $ as explained in subsection \ref{Wilson}.
For the identity block, we set $S$ as the direct product of two singlets given by $s_{12} \in \bar{\Lambda} \otimes \Lambda $ and $s_{34} \in \bar{\Xi} \otimes \Xi$. With the choice, the Wilson line network factorizes as
\begin{align}
G_4 (\Lambda_j | z_j) = \left( \langle s_{12}| \prod_{j=1,2} W_{\Lambda_j} (z_0;z_j) | \text{hw} \rangle_j \right) \left( \langle s_{34}| \prod_{l=3,4} W_{\Lambda_l} (z_0' ;z_l) | \text{hw} \rangle_l \right) 
\end{align}
with $z_0 = z_0'$. Setting $z_0 = z_1 = \infty$ and $z_0' = z_3 = z$ with $z_2 =1$ and $z_4=0$, the Wilson line network becomes the product of two open Wilson lines. 
Therefore, the identity block should be computed as
\begin{align}
\langle G_4 (\Lambda_j | z_j)  \rangle = \langle W_{\Lambda} (\infty ; 1) W_{\Xi} (z; 0) \rangle \, ,
\label{GW4}
\end{align}
where the products of currents are evaluated by the correlators of W$_N$ theory as before. 
In this subsection we explicitly compute \eqref{GW4} for $N=3$ with $\Lambda = (\lambda_1,\lambda_2)$ and $\Xi = \text{f}$, that is
\begin{align}
G^0_4(z)= \langle W_{(\lambda_1 , \lambda_2)} (\infty ; 1) W_{\text{f}} (z; 0) \rangle
= \lim_{w \to \infty} w^{2 h(\lambda_1,\lambda_2;0)} \langle W_{(\lambda_1 , \lambda_2)} (w ; 1) W_{\text{f}} (z; 0) \rangle \, ,
\label{4ptWilson}
\end{align}
by applying the method developed in the previous section.
We compare the result with the identity block from the CFT computation in \eqref{coulomb}, see appendix A for some details. The $1/c$ expansion is given as \eqref{1/cexp0} with \eqref{1/cexp}, but now we set $N=3$ and $m=1$.

\begin{figure}
	\centering
	\includegraphics[keepaspectratio, scale=0.49]
	{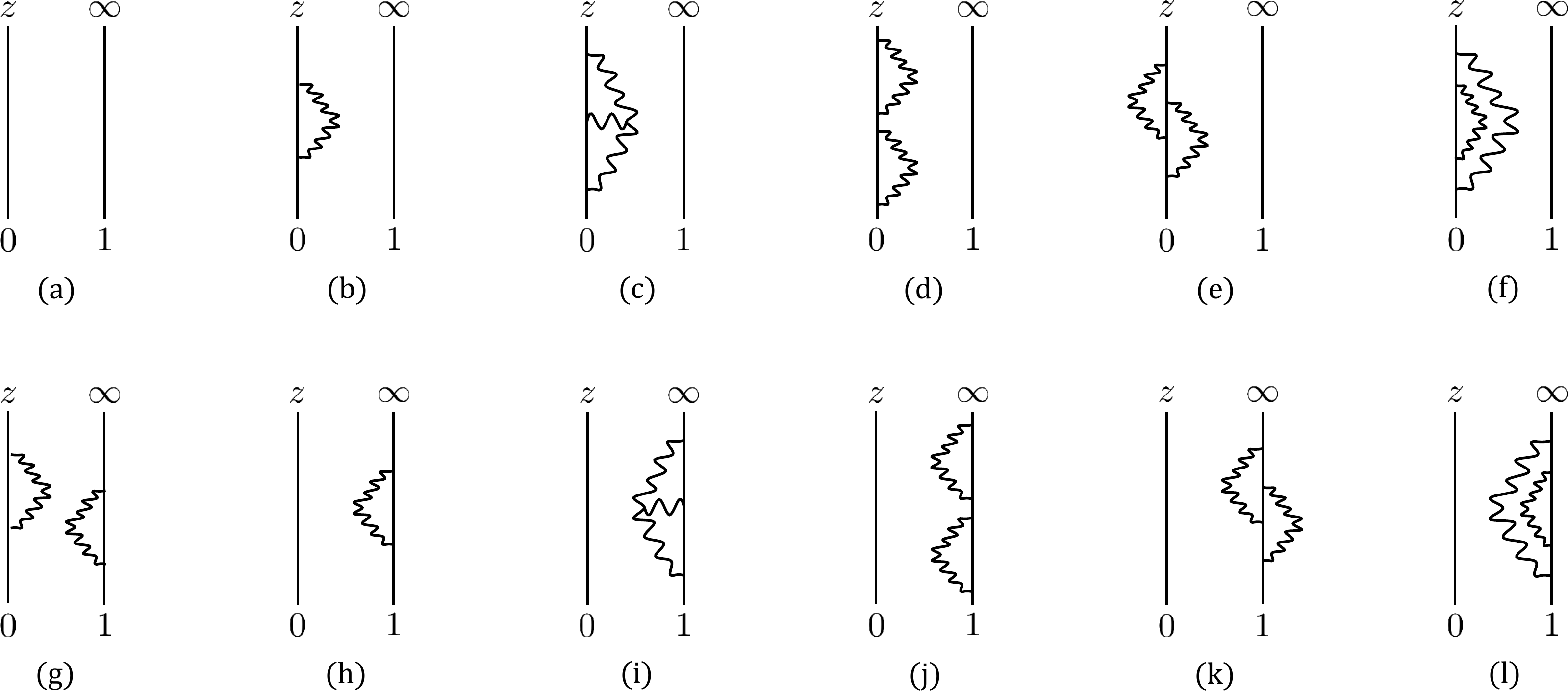}
	\caption{Contribution to the identity W$_3$ block from the product of open Wilson lines at the leading order in $1/c$ and its corrections of self-energy type up to the order $1/c^2$. The straight and wavy lines represent open Wilson lines and propagators of current with spin $s=2$ or $s=3$. (a) Leading order contribution. (b)-(f) Corrections associated with the open Wilson line from 0 to $z$. (g) Correction associated with both the open Wilson lines. (h)-(l)  Corrections associated with the open Wilson line from 1 to $\infty$.}
	\label{CBWilson1}
\end{figure}

The leading order expression in $1/c$ is simply given by the product of two expectation values as
\begin{align}
	\left. G_4^{0}(z) \right |_{\mathcal{O}(c^0)}   =  \left. \langle W_{(\lambda_1 , \lambda_2)} (\infty ; 1) \rangle \right|_{\mathcal{O}(c^0)} \left.
	\langle W_{\text{f}} (z; 0) \rangle \right |_{\mathcal{O}(c^0)}  = z^2 \, . \label{G4Wtree}
\end{align}
This can be expressed in diagram (a) of figure \ref{CBWilson1}.
Among higher order corrections in $1/c$, there are contributions of the self-energy type as in other diagrams.
For contributions corresponding to diagrams (b)-(f), we replace $\left.
\langle W_{\text{f}} (z; 0) \rangle \right |_{\mathcal{O}(c^0)} $ in \eqref{G4Wtree} the integrals of the forms
\begin{align}
\int_0^z dz_n \cdots \int_0^{z_2} dz_1 f^{(s_n, \ldots , s_1)}_{\text{f}} (z;0;z_n , \ldots,z_1)
 \langle  J^{(s_n)}(z_n) \cdots J^{(s_1)} (z_1) \rangle
\end{align}
with $n=2,3,4$ and $s_j = 2,3$. Here, the integrals with $n=2$ and $n=3$ correspond to diagrams (b) and (c). The integrals with $n=4$ include the four point functions of currents with several different terms as in \eqref{4ptspins} and \eqref{4ptcurrent}, which are expressed by diagrams (d), (e), and (f).
The $1/c$ corrections to $W_{\text{f}} (z; 0) $ up to order $1/c^2$ have been evaluated in \cite{Hikida:2017ehf}, and
they can be summarized as the shift of conformal dimension from \eqref{G4Wtree} to (see \eqref{hexp} for a general operator)
\begin{align}
\frac{1}{z^{2 h(\text{f};0)}} \, , \quad h(\text{f};0) = - 1 - \frac{32}{c} - \frac{1600}{c^2} + \mathcal{O}(c^{-3})
\label{VBself}
\end{align}
after renormalizing the overall factor of the open Wilson line and the coupling constants $c_s$ in \eqref{ren}.
The other diagrams correspond to the similar corrections of another Wilson line $W_{(\lambda_1 , \lambda_2)} (\infty ; 1) $, but they do not appear in the four point block $G_4^0(z)$ \eqref{4ptWilson} since an operator is put at the infinity.

\begin{figure}
	\centering
	\includegraphics[keepaspectratio, scale=0.49]
	{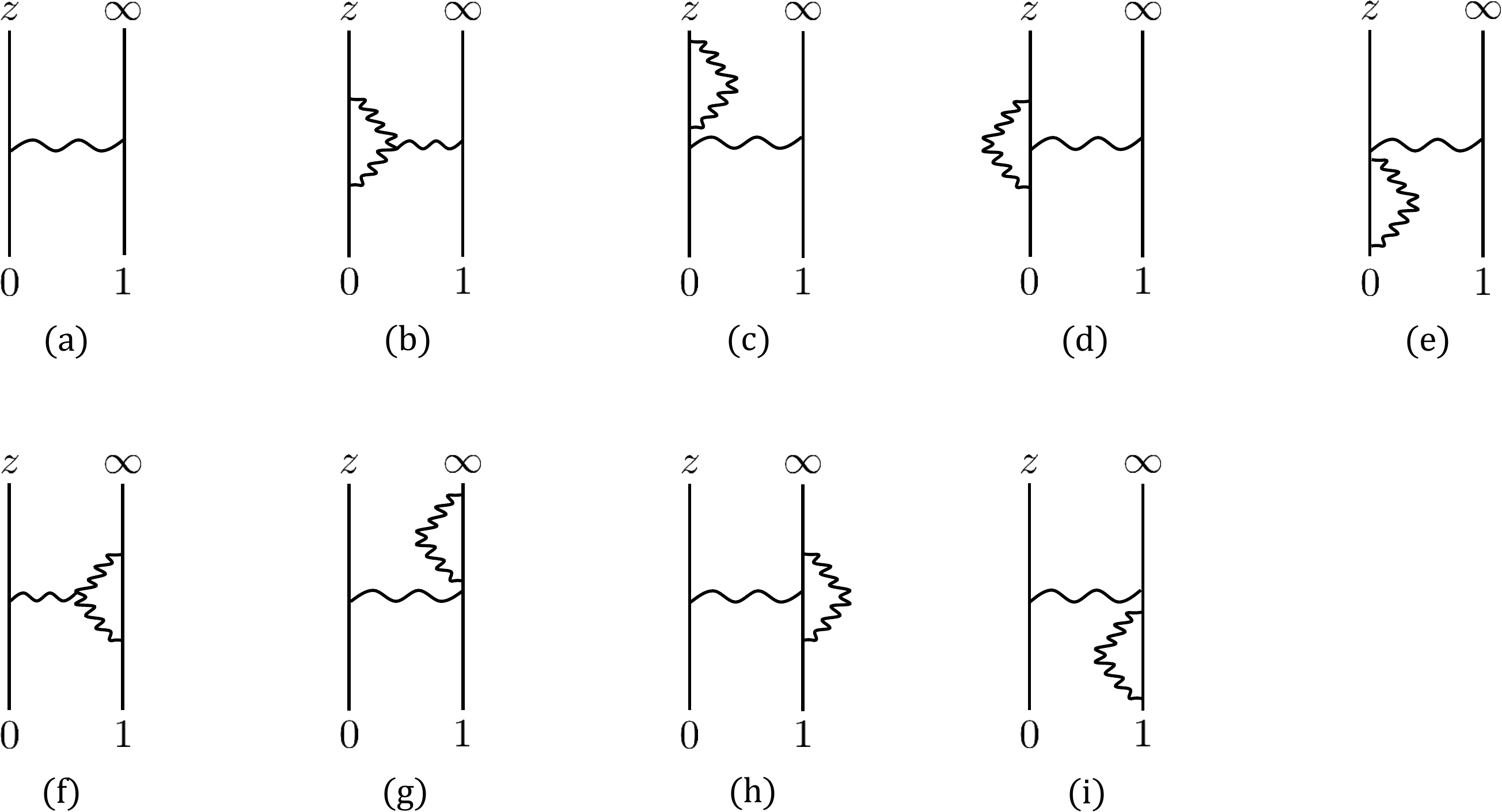}
	\caption{(a) Contribution with the exchange of a current with $s=2$ or $s=3$. (b)-(i) Its $1/c$ corrections from loop diagrams.}
	\label{CBWilson2}
\end{figure}

Non-trivial contributions start to enter from the next order in $1/c$.
They correspond to diagram (a) of figure \ref{CBWilson2},
and the sum of them is
\begin{align}
 H_1 (z) = H_1^{(2)} (z) + H_1^{(3)} (z)
 \label{H1}
\end{align}
with
\begin{align}
	H_1^{(s)} (z) = \left( \frac{6}{c}\right)^2  \int_1^\infty dz_2 \int_0^{z} dz_1 	f^{(s)}_{(\lambda_1, \lambda_2)} (\infty;1;z_2) f_{\text{f}}^{(s)} (z;0;z_1)
 \langle J^{(s)} (z_2) J^{(s)} (z_1) \rangle \, .
\end{align}
With the two point functions of currents in \eqref{2ptcurrent} with \eqref{B2B3}, we evaluate the integrals as
\begin{align}
	H_1^{(2)}(z) =  \frac{2 (\lambda_1 + \lambda_2)}{c} z^4 {}_2 F_1(2,2;4;z) \, , \quad
	H_1^{(3)} (z)= - \frac{2(\lambda_1 - \lambda_2)}{15 c} z^5 {}_2 F_1(3,3;6;z) \, .
	\label{G123}
\end{align}
They correspond to the global blocks with spin 2 and 3 currents as the intermediate operators as they should be.
The coefficients are also given by the products of three point functions divided by a current-current two point function as
\begin{align}
	H_1^{(2)} (z) =\frac{h_{(\text{f};0)} h_{(\lambda_1 , \lambda_2 ; 0)}}{B_2} z^4 {}_2 F_1 (2,2;4;z) \, , \quad
	H_1^{(3)} (z)= \frac{ w_{(\text{f};0)} w_{(\lambda_1 , \lambda_2 ; 0)}}{B_3} z^5 {}_2 F_1 (3,3;6;z)
\end{align}
at the leading order in $1/c$.
Here, the spin 2 and 3 charges were given by \eqref{hexp} and \eqref{wexp}.

The other diagrams in figure \ref{CBWilson2} yield loop corrections to interactions among an open Wilson line and an exchanged current. These diagrams can be evaluated precisely as for \eqref{G3s} but with $y$ replaced by $z_1$ or $z_2$. Therefore, after renormalizing the open Wilson lines and the coupling constants $c_s$, the net effects are summarized as the $1/c$ order terms in \eqref{hexp} and \eqref{wexp} as
\begin{align}
\begin{aligned}
	&h_{(\text{f};0)} h_{(\lambda_1 , \lambda_2 ; 0)} =  \lambda_1 + \lambda_2
	+ \frac{8 (7 \lambda_1 + \lambda_1^2 + 7 \lambda_2  + \lambda_2^2 + \lambda_1 \lambda_2)}{c} + \mathcal{O}(c^{-2}) \, , \\
	& w_{(\text{f};0)} w_{(\lambda_1 , \lambda_2 ; 0)} = \frac{\lambda_1-\lambda_2}{9} + \frac{8 \left(15 \lambda_1^2+41 \lambda_1-15 \lambda_2^2-41 \lambda_2\right)}{45 c}+ \mathcal{O}(c^{-2}) \, .
\end{aligned}
\end{align}
Thus we have a type of $1/c^2$ corrections as
\begin{align}
\begin{aligned}
	&H_2^{(2)} (z)=   \frac{16 (7 \lambda_1 + \lambda_1^2 + 7 \lambda_2  + \lambda_2^2 + \lambda_1 \lambda_2)}{c^2} z^4 F(2,2,4;z) \, , \label{G223}\\
	&H_2^{(3)} (z)= - \frac{16 \left(15 \lambda_1^2+41 \lambda_1-15 \lambda_2^2-41 \lambda_2\right)}{75 c^2} z^5 F(3,3,6;z)
\end{aligned}
\end{align}
coming from diagrams (b)-(i) of figure \ref{CBWilson2}.

\begin{figure}
	\centering
	\includegraphics[keepaspectratio, scale=0.49]
	{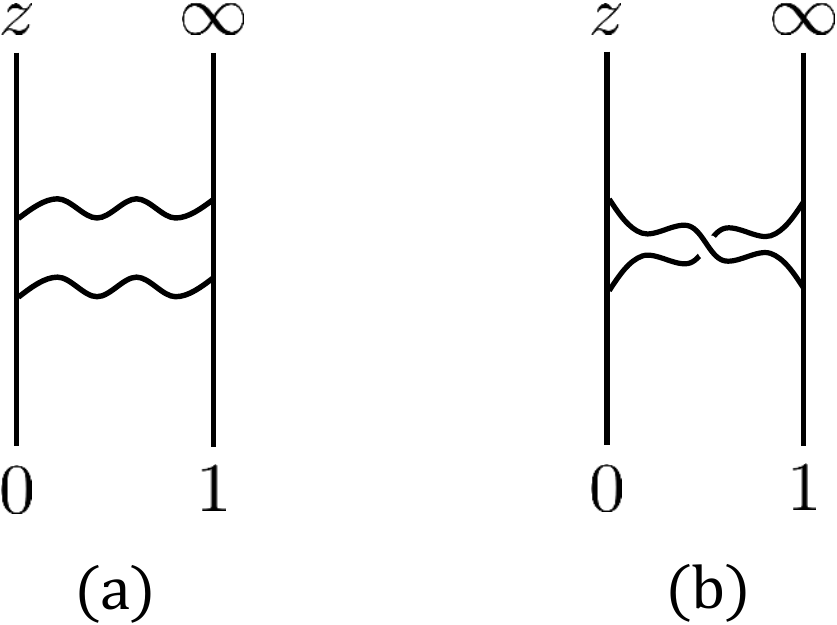}
	\caption{(a),(b) Contributions with the exchange of two currents with $s=2$ or $s=3$.}
	\label{CBWilson3}
\end{figure}

At the order of $1/c^2$, there are more interesting contributions with the exchange of two currents as in diagrams (a) and (b) of figure \ref{CBWilson3} in addition to those already given.
When the two currents have the same spin, they come from the integrals
\begin{align}
\begin{aligned}
	H^{(s,s)}_2 (z)& = \left( \frac{6}{c}\right)^4  B_s^2 \int_1^\infty dz_4 \int_1^{z_4} dz_3 \int_0^z dz_2 \int_0^{z_2} dz_1  \\ &\times
	f^{(s,s)}_{(\lambda_1, \lambda_2)} (\infty;1;z_4,z_3)	f_{\text{f}}^{(s,s)} (z;0;z_2,z_1)
	\left (\frac{1}{z_{31}^{2s} z_{42}^{2s}} + \frac{1}{z_{41}^{2s} z_{32}^{2s}} \right)
\end{aligned}
\end{align}
with $s=2,3$. The first and second terms are expressed by diagrams (a) and (b), respectively.
When the two currents have the different spins, the contributions corresponding to diagrams (a) and (b) are given by the integrals
\begin{align}
\begin{aligned}
	H^{(s_1,s_2 )}_{2,1} (z)& = \left( \frac{6}{c}\right)^4  B_2 B_3 \int_1^\infty dz_4 \int_1^{z_4} dz_3 \int_0^z dz_2 \int_0^{z_2} dz_1  \\ &\times
	f^{(s_2,s_1)}_{(\lambda_1, \lambda_2)} (\infty;1;z_4,z_3) 
	f_{\text{f}}^{(s_2,s_1)} (z;0;z_2,z_1)
\frac{1}{z_{31}^{2s_1} z_{42}^{2s_2}} \, , \\
	H^{(s_1,s_2 )}_{2,2} (z)& =  \left( \frac{6}{c}\right)^4  B_2 B_3 \int_1^\infty dz_4 \int_1^{z_4} dz_3 \int_0^z dz_2 \int_0^{z_2} dz_1  \\ &\times
	f^{(s_1,s_2)}_{(\lambda_1, \lambda_2)} (\infty;1;z_4,z_3) 
	f_{\text{f}}^{(s_2,s_1)} (z;0;z_2,z_1)
\frac{1}{z_{41}^{2s_1} z_{32}^{2s_2}}
\end{aligned}
\end{align}
with $(s_1,s_2) = (2,3)$ or $(3,2)$.
Summing over this type of contributions, we find
\begin{align}
	&H_2^\text{2ex} (z) \nonumber \\
	&= \frac{8}{5} \bigl[20 \log ^2(1-z) (3 (\lambda_1^2+\lambda_1 (4 \lambda_2+6)+(\lambda_2-3) \lambda_2)+z^2 (2 \lambda_1+\lambda_2) (2 \lambda_1+\lambda_2+9) \nonumber\\ & -6 \lambda_1 z (\lambda_1+2 \lambda_2+6))+3 z (10 \lambda_1^2 (31 z-30)+\lambda_1 (160 \lambda_2 z+519 z-318)  \nonumber \\ & +\lambda_2 (10 \lambda_2 (z+30)+201 z+318))-6 \log (1-z) (10 \lambda_1^2 (8 (z-3) z+15) \label{G2main}\\ & +\lambda_1 (2 z (30 \lambda_2 (z-2)+7 z+3)-21)+\lambda_2 (10 \lambda_2 (z (z+6)-15)+(z-36) z+21) )\nonumber \\ & +360 \text{Li}_2(z) (\lambda_1 (2 (z-3) z+3)+\lambda_2 (z^2-3) ) \bigr] \nonumber
\end{align}
with $\text{Li}_2 (z) = \sum_{n=1}^\infty z^n/n^2$.
The total contribution at the $1/c^2$ order other than the self-energy type is thus
\begin{align}
	H_2 (z)= H_2^{(2)} (z) + H_2^{(3)} (z) + H_2^\text{2ex} (z)
	\label{G2total}
\end{align}
with \eqref{G223} and \eqref{G2main}.

We confirm the bulk computation by showing that the above result reproduces the CFT one. The identity block in \eqref{coulomb} is expanded in $1/c$ as \eqref{1/cexp0} with \eqref{1/cexp}, and we set $N=3$, $m=1$ and identify $(\lambda_1,\lambda_2)=(n,0)$ for the comparison.
We can see that  the $1/c$ order term in \eqref{1/cexp0} is the same as \eqref{H1} with \eqref{G123}.
Furthermore. we can check that the $1/c^2$ order expression of \eqref{1/cexp0} with \eqref{1/cexp} is reproduced by \eqref{G2total} at each order in $z$ expansion.

Before ending this subsection, we would like to mention the relation to the analysis in \cite{Fitzpatrick:2016mtp} for $N=2$ with a different regularization. The regularization prescription in \cite{Fitzpatrick:2016mtp} is to adopt the normal ordered product for currents coming from the same Wilson line operator. Moreover, they used the exact conformal weight $h_j$ from the beginning instead of the quantum number $j$ for the finite representation of sl$(2)$. Using the normal ordered prescription is equivalent to neglecting contributions coming from loop diagrams in figures \ref{CBWilson1} and \ref{CBWilson2}, see also \cite{Besken:2017fsj}. As we explained above, these contributions only shift the conformal weight for $N=2$; therefore, the effect is already included in their treatment. The contributions represented by diagrams in figure \ref{CBWilson3} are the same in the both prescriptions, see figure 5 of \cite{Fitzpatrick:2016mtp}. Therefore, we can see that the both ways of computation should give the same result for $N=2$.

\subsection{General conformal blocks}
\label{GenericCB}

In this subsection, we study the conformal blocks with the exchange of the general operator from the networks of open Wilson lines \eqref{WLN} by including quantum effects.
For the general blocks, it turns out to be convenient to use the $X$-basis as
\begin{align}
G_n(\Lambda_j | z_j) =\left[  \prod_{j = 1}^n \int d X_j  \,  {}_{j} \langle  \text{lw} | X_j \rangle  W_{\Lambda_j} (z_j  ; z_0 ; X_j )   \langle X_j|  \right] | S \rangle
\end{align}
in its conjugated form. Here the generators $V_m^{s}(X_j)$ are now written in terms of $X_j$ 
as explicitly indicated in the arguments of the Wilson line operator.
The condition for the singlet $|S \rangle$ becomes
\begin{align}
\sum_{j=1}^n  V_m^{s} (X_j)  \left[ \prod_{l=1}^n \langle X_l | \right] | S \rangle  = 0 
\label{singlet}
\end{align}
for all $s$ and $m$. 
In general there are several solutions, and a solution corresponds to a conformal block.
Applying $ {}_{j} \langle  \text{lw} | X_j \rangle  = \delta (X_j)$ as in \eqref{statesN2} or \eqref{Xlw}, the expression is simplified as
\begin{align}
G_n(\Lambda_j | z_j) = \left. \left[  \prod_{j = 1}^n   W_{\Lambda_j} (z_j  ; z_0 ; X_j )   \langle X_j|  \right] | S \rangle  \right|_{X_j = 0} \, .
\label{GnW}
\end{align}
The quantum effects are evaluated by using the correlators of higher spin currents inserted as before.

\begin{figure}
	\centering
	\includegraphics[keepaspectratio, scale=0.49]
	{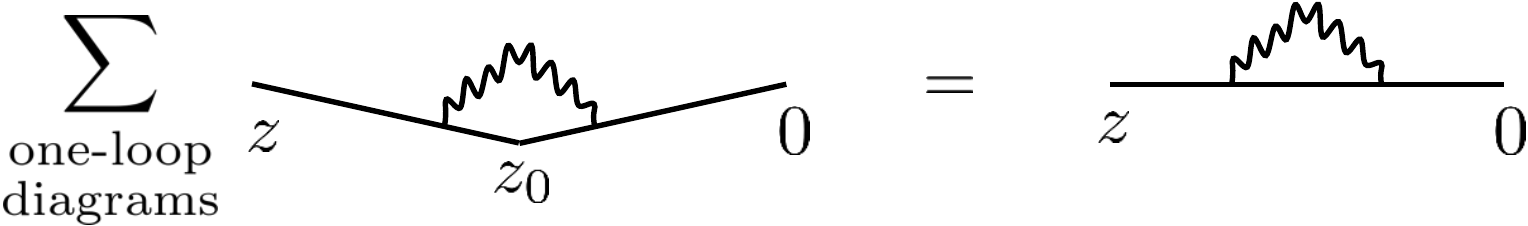}
	\caption{Contribution to the two point function from the network of open Wilson lines at the $1/c$ order.}
	\label{3ptWilson0}
\end{figure}

In the following we mainly focus on the $N=2$ case and make some comments on the $N=3$ case at the end of this subsection.
Before going into the details on three and four point Virasoro blocks, let us mention how the formalism explained above reproduces the previous analysis on the two point function for $N=2$.
We examine $G_2 (j,j|z,0)$ in \eqref{GnW} with $X_i = x_i$.
The vertex is given by the unique solution to the singlet condition \eqref{singlet} as (see, e.g., \cite{Fateev:2011qa})
\begin{align}
 \langle x_1 |   \langle x_2|| S \rangle = v_2(j , j | x_1 , x_2)  \equiv   x_{12}^{2 j} 
\label{g2ptx}
\end{align}
up to the overall factor. 
At the leading order in $1/c$, we have
\begin{align}
\left. \langle G_2 (j , j | z , 0) \rangle \right|_{\mathcal{O}(c^0)} = \left.  e ^{(z - z_{0})V_1^{2 }(x_1) } e ^{ - z_{0}V_1^{2 }(x_2) }
v_2 (j , j | x_1 , x_2) \right|_{x_i = 0}
= z^{2 j} 
\label{g2ptz}
\end{align}
as in \eqref{2ptN2tree}. For the $1/c$ corrections, it is convenient to set $z_0 = 0$ (or $z_0 =z$), which reduces the number of diagrams, see figure \ref{3ptWilson0}.
With $z_0 = 0$, the expression in \eqref{GnW} reduces to the expectation value of \eqref{G2x} with \eqref{statesN2} and $(z_f,z_i) = (z,0)$. We have also evaluated the $1/c$ corrections to the two point function with keeping $z_0$ generic and checked that the final result is the same.

\subsubsection{Three point Virasoro block}

As a simple but non-trivial example, we study three point conformal block for $N=2$. The three point block is fixed by the conformal symmetry as
\begin{align}
\langle \mathcal{O}_{j_1} (z_1) \mathcal{O}_{j_2} (z_2) \mathcal{O}_{j_3} (z_3) \rangle
\propto \frac{1}{z_{12}^{h_{j_1} + h_{j_2} - h_{j_3}} z_{13}^{h_{j_1} + h_{j_3} - h_{j_2}} z_{32}^{h_{j_3} + h_{j_2} - h_{j_1}}}
\label{3ptscalar}
\end{align}
with (see \eqref{hN2})
\begin{align}
h_{j_i} = - j_i +  \frac{1}{c} h_{i}^{(1)} + \mathcal{O}(c^{-2}) \, , \quad h_{i}^{(1)} = - 6 j_i (j_i + 1) \, . \label{hi}
\end{align}
%We shall reproduce this form from the network of open Wilson lines.
We evaluate $G_3 (j_i | z_i)$ in  \eqref{GnW} with $X_i = x_i$.
The vertex is uniquely fixed by the singlet condition \eqref{singlet} as  (see, e.g., \cite{Fateev:2011qa,Fitzpatrick:2016mtp})
\begin{align}
\left[ \prod_{i=1}^3 \langle x_i | \right] | S \rangle = v_3(j_i | x_i)  \equiv   x_{12}^{j_1 + j_2 - j_3} x_{13}^{j_1 + j_3 - j_2} x_{32}^{j_3 + j_2 - j_1} 
\label{g3ptx}
\end{align}
up to the overall factor.
At the leading order in $1/c$, we find
\begin{align}
\begin{aligned}
\left. \langle G_3 (j_i | z_i) \rangle \right|_{\mathcal{O}(c^0)} = \left. \left[ \prod_{j=1}^3 e ^{z_{j0}V_1^{2 }(x_j) } \right]
v_3(j_i | x_i) \right|_{x_i = 0}
= z_{12}^{j_1 + j_2 - j_3} z_{13}^{j_1 + j_3 - j_2} z_{32}^{j_3 + j_2 - j_1} \, .
\end{aligned}
\label{g3ptz}
\end{align}
The expression reproduces the three point function at the leading order in $1/c$.
Here we simply replace $x_i$ by $z_i$,%
\footnote{Precisely speaking, $x_i$ are replaced by $z_{i0} \, (\equiv z_i -z_0)$ with $z_0$ as the position of vertex for the Wilson line network. Since the results do not depend on $z_0$, we can set $z_0 = 0$.}
and this is actually as the same as the prescription to obtain the large $c$ limit of three point Virasoro block in \cite{Fateev:2011qa}.

\begin{figure}
	\centering
	\includegraphics[keepaspectratio, scale=0.49]
	{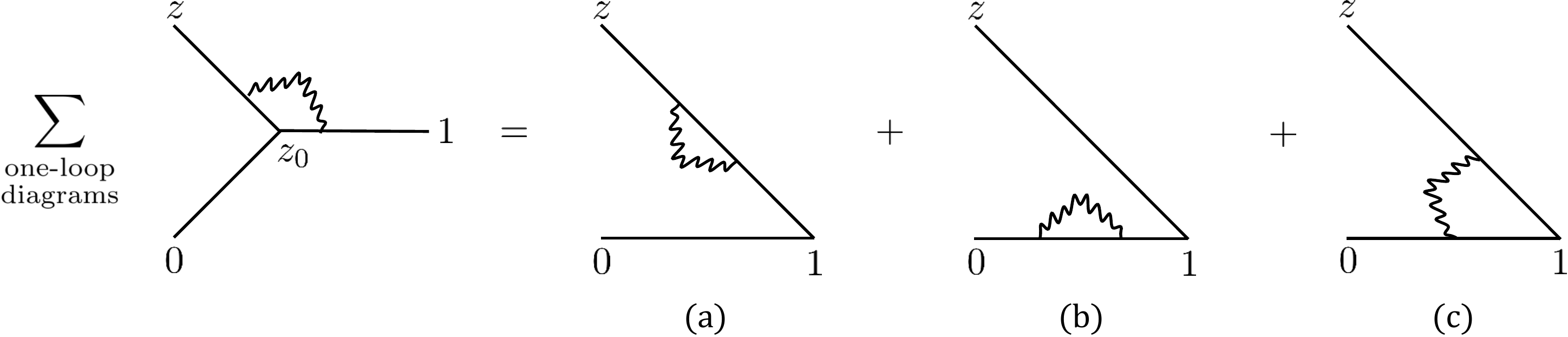}
	\caption{Contributions to the three point function from the network of open Wilson lines at the $1/c$ order. (a),(b) Corrections of self-energy type. (c) Correction from the exchange of a current with $s=2$.}
	\label{3ptWilson1}
\end{figure}

We move to the next leading order in $1/c$.
When evaluating the expectation value of $G_3 (j_i | z_i)$, the correlators of $J^{(2)} (z)$ for Virasoro algebra are used, and divergences are removed as before.
In order to make the computation simpler, we set $(z_1,z_2,z_3) = (z,0,1)$.
Moreover, using $z_0 = z_3 = 1$,%
\footnote{Similarly to the two point function, we have checked that the final result in \eqref{G3final} does not depend on $z_0$ even for generic $z_0$. For more complicated correlators, the computations would be quite messy without setting $z_0$ as a specific value.}
we consider the product of two Wilson lines as in figure \ref{3ptWilson1}.
With this setup, we evaluate the integrals corresponding to the three diagrams in the right hand side.
Defining
\begin{align}
&t_{1} ( z ; x_1 , x_2 , x_3) =  \left( \frac{6}{c} \right)^2  \int_{1}^z dz_2 \int_1^{z_2} d z_1
\prod_{j=1}^2 \left[ e^{(z - z_j) V_1^{2} (x_1)} V_{- 1}^{2} (x_1) e^{(z_j - z) V_1^{2 } (x_1)}  \right]
\nonumber \\
& \quad \times
e^{(z-1) V_1^{2} (x_1)} e^{-V_1^{2} (x_2)} v_3(j_i | x_i)
\langle J^{(2)}(z_2) J^{(2)}(z_1) \rangle \, ,  \nonumber \\
&t_{2}  ( z ; x_1 , x_2 , x_3)  =  \left( \frac{6}{c} \right)^2 \int_1^0 dz_2 \int_1^{z_2} d z_1
\prod_{j=1}^2 \left[ e^{ - z_j V_1^{2} (x_2)} V_{- 1}^{2} (x_2) e^{ z_j V_1^{2} (x_2)}  \right]         \nonumber \\
& \quad \times
e^{(z-1) V_1^{2} (x_1)} e^{-V_1^{2} (x_2)} v_3(j_i | x_i)
\langle J^{(2)}(z_2) J^{(2)}(z_1) \rangle \, , \label{ta}  \\
& t_{3}  (z ; x_1 , x_2 , x_3) = \left( \frac{6}{c} \right)^2 \int_1^0 dz_2 \int_1^{z} d z_1
\left[ e^{ - z_2 V_1^{(2)} (x_2)} V_{- 1}^{(s) } (x_2) e^{ z_2 V_1^{(2) } (x_2)}  \right]  \nonumber
\\
& \quad \times
\left[ e^{(z - z_1) V_1^{2} (x_1)} V_{- 1}^{2} (x_1) e^{(z_1 - z) V_1^{2} (x_1)}  \right]  e^{(z-1) V_1^{2} (x_1)} e^{-V_1^{2} (x_2)} v_3(j_i | x_i)
\langle J^{(2)}(z_2) J^{(2)}(z_1) \rangle \, , \nonumber
\end{align}
the integrals corresponding to diagrams (a), (b), and (c) are
$ t_a (z ; 0,0,0) $ with $a=1,2,3$, respectively.
The $1/c$ order correction to the three point conformal block is given by the sum of them as
\begin{align}
\begin{aligned}
\left. \langle G_3 (j_i | z_i) \rangle \right|_{\mathcal{O}(c^{-1})} = &\sum_{a=1}^3 t_a (z ; 0,0,0) =
- \frac{1}{c} z^{j_1+j_2-j_3}  (z-1)^{j_1+j_3-j_2} \\
 &  \times  \left[ (h^{(1)}_1 + h^{(1)}_2 - h^{(1)}_3) \log(z) + (h^{(1)}_1 + h^{(1)}_3 - h^{(1)}_2) \log (z-1) \right]
 \end{aligned}
 \label{G3final}
\end{align}
with $h_i^{(1)}$ in \eqref{hi} up to the order $\epsilon^0$ term. Here we have ignored the term proportional to the tree level expression, which is interpreted as a correction to the overall factor of three point function.
The result reproduces the $1/c$ order correction of three point block in \eqref{3ptscalar} with \eqref{hi} by setting $(z_1,z_2,z_3) = (z,0,1)$.

\subsubsection{Four point Virasoro block}
\label{g4pt}

We move to the Virasoro block of operator $\mathcal{O}_p$ for the four point function \eqref{4ptN2}. For this correlator, we may set $(z_1,z_2,z_3,z_4) = (z,0,\infty,1)$ and $(j_1,j_2,j_3,j_4) = (j,j,q,q)$.
We compute $G_4(j_i | z_i)$ in \eqref{GnW} with $X_i = x_i$.
A solution to the singlet condition \eqref{singlet} is given by \cite{Fateev:2011qa}
\begin{align}
\left[ \prod_{i=1}^4 \langle x_i | \right] | S \rangle = v_p(j_i | x_i) \, ,
\end{align}
where we have defined 
\begin{align}
v_p (j_i | x_i) \equiv \mathcal{C} \int d x \, v_3 (j ,j , p | x_1 ,x_2 ,x) v_3 (q,q , -1 -p|  x_3 ,x_4 , x) \, .
\label{g4ptx}
\end{align}
With the integral contour as $x \in (0 , \infty)$ and the overall factor as
\begin{align}
\mathcal{C} = (-1)^p \frac{\Gamma ( -2 p)}{(\Gamma (-p))^2 } \, ,
\end{align}
the leading order expression in $1/c$ becomes
\begin{align}
\left. \langle G_4(j_i |  z_i) \rangle \right|_{\mathcal{O}(c^0)} = \left. \left[ \prod_{j=1}^4 e ^{z_{j0}V_1^{2 }(x_j) } \right]
v_p (j_i | x_i) \right|_{x_i = 0} = z^{2j - p} g_p (z) \, ,
\label{g4ptz}
\end{align}
where $z^{2j - p} g_p (z)$ is the  global block of operator $\mathcal{O}_p$ as
\begin{align}
z^{2j - p} g_p (z)  \equiv z^{2j - p} {}_2 F_1 (-p,-p;-2p;z) \, .
\label{globalN2}
\end{align}
Just like the  three point block, this is essentially the way to obtain the large $c$ limit of the four point block in \cite{Fateev:2011qa}.

\begin{figure}
	\centering
	\includegraphics[keepaspectratio, scale=0.49]
	{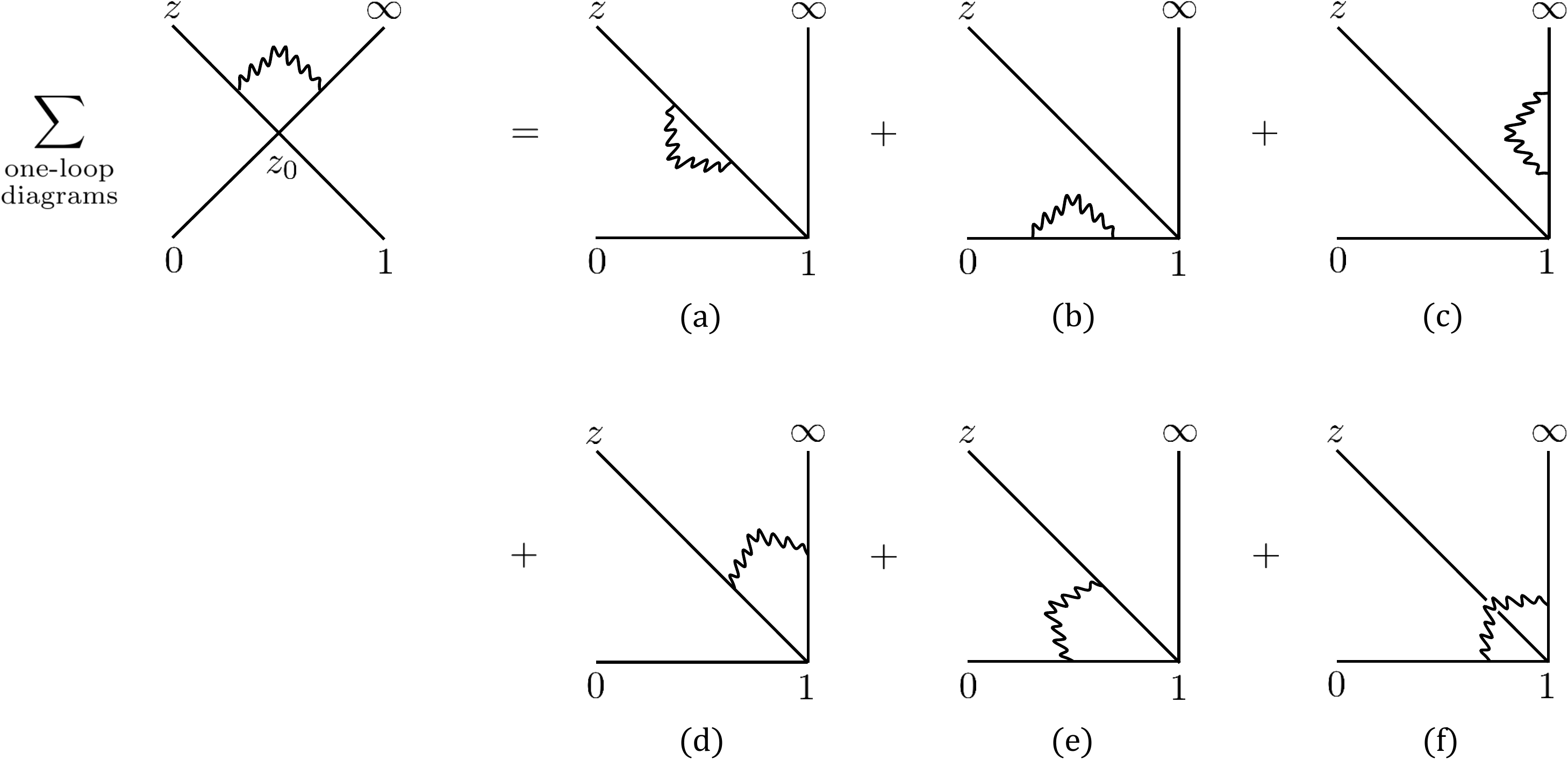}
	\caption{Contributions to the four point Virasoro block of operator $\mathcal{O}_p$ from the network of open Wilson lines at the $1/c$ order. (a)-(c) Corrections of self-energy type. (d)-(f) Corrections from the exchange of a current with $s=2$.}
	\label{3ptWilson2}
\end{figure}

For the next leading order in $1/c$, we evaluate the network of open Wilson lines with $z_0 = z_4 = 1$. With this setting, the number of diagrams we have to consider is reduced as in figure \ref{3ptWilson2}.
We choose the normalization of conformal block as
\begin{align}
\langle G_4(j_i | z_i) \rangle  = z^{2j - p} (1 + \mathcal{O}(z) ) \, ,
\end{align}
and hence we subtract the terms proportional to the global block \eqref{globalN2} such that there is no $z^{2j -p}$ term at the $1/c$ order.
With $t_a (z;x_1,x_2,x_3)$ defined in \eqref{ta},
the integrals corresponding to diagrams (a), (b), and (e) are written as
\begin{align}
\mathcal{C}   \int_{0}^\infty d x \,  t_a ( z ; 0,0,x) x^{ - 1 - p}
\end{align}
with $a=1,2,3$, respectively.
For other diagrams, we define the functions as
\begin{align}
u_{1} (z ; z' ;  x_1 , x_2 , x_3) =&  \left[ e^{(z - z') V_1^{2} (x_1)} V_{- 1}^{2} (x_1) e^{(z' - z) V_1^{2} (x_1)} \right]
e^{(z - 1) V_1^{2} (x_1)} e^{ - V_1^{2} (x_2)} v_3 (j_i | x_i)\, , \nonumber \\
u_{2} (z ; z' ; x_1 , x_2 , x_3) =&  \left[  e^{ - z' V_1^{2} (x_1)} V_{- 1}^{2} (x_2) e^{ z' V_1^{2} (x_1)}\right] e^{(z - 1) V_1^{2} (x_1)} e^{ - V_1^{2} (x_2)} v_3 (j_i | x_i)
\end{align}
with $j_1 = j_2 = j, j_3 = p$, and
\begin{align}
\begin{aligned}
u_{3} (z' ;  x_1 , x_2 , x_3)   = & \lim_{w \to \infty} w^{-2 q} \left[e^{(w - z') V_1^{2} (x_1)} V_{- 1}^{2} (x_1) e^{(z' - w ) V_1^{2 } (x_1)}  \right]
e^{(w - 1) V_1^{2} (x_1)} v_3 (j_i | x_i)
\end{aligned}
\end{align}
with $j_1 = j_2 = q , j_3 = - 1 - p$.
Then the integrals corresponding to diagrams (d) and (f) are
\begin{align}
\begin{aligned}
&\mathcal{C} \left(\frac{6}{c} \right)^2  \int_1^z dz_2 \int_1^\infty dz_1 \int_0^\infty dx \, u_1(z ;z_2 ; 0,0,x) u_3 (z_1 ;0,0,x)  \langle  J^{(2)} (z_1) J^{(2)} (z_2) \rangle \, , \\
&\mathcal{C} \left(\frac{6}{c} \right)^2\int_1^0 dz_2 \int_1^\infty dz_1 \int_0^\infty d x \, u_2(z ;z_2 ; 0,0,x) u_3 (z_1;0,0,x)  \langle J^{(2)} (z_1) J^{(2)} (z_2) \rangle \, ,
\end{aligned}
\end{align}
respectively.
The sum of all contributions is computed as%
\begin{align}
\begin{aligned}
 &	\left. \langle G_4 (j_i | z_i) \rangle \right|_{\mathcal{O}(c^{-1})} = 
 \frac{6  z^{2 j-p-1}}{c} \Bigl\{ {}_2F_1(-p,-p;-2 p;z)  \bigl[ z \log (z) (2 j (j+1)- p (p+1))
 \\
& \qquad + \log (1-z) \left(2 j q (z-2)+p^2 (z-1)+p (z-1) \right)
+z (2 j-p) (p-2 q)\bigr]    \\ &  \qquad  -  {}_2F_1(1-p,-p;-2 p;z) \log (1-z)  p (z-1) (2 j+2 q+1)  \Bigr\}
\end{aligned}
\label{resultN2}
\end{align}
up to the terms proportional to \eqref{globalN2}.
Here we neglected the integral corresponding to diagram (c) since it can be shown to give only terms proportional to \eqref{globalN2} or $h_p \log(w)$ with $w \to \infty$.

Let us compare the $1/c$ order expression with the known result in \eqref{Vp}.
First we examine the $\log (z)$ term in \eqref{resultN2}.
This type of contribution comes from the shifts of conformal weights as
\begin{align}
\begin{aligned}
&z^{ - 2 h_j  + h_p} {}_2 F_1 (h_p ,h_p ;2 h_p;z) \\
&\quad =  z^{2j - p} g_p (z)
+ \frac{1}{c} (- 2 h^{(1)}_j + h^{(1)}_p) \log(z)   z^{2j - p} g_p (z)+ \frac{1}{c} z^{2j - p} r_p (z) + \mathcal{O}(c^{-2})
\end{aligned}
\end{align}
with
\begin{align}
r_p (z) = - 12 p (p+1)  \sum _{n=0}^{\infty }\frac{\left[ (-p)_n\right]^2}{ (-2 p)_n} \frac{z^n }{n!} \sum _{k=1}^n \left(\frac{1}{k-p-1}-\frac{1}{k-2 p-1}\right) \, .
\end{align}
We can see that the term proportional to $\log (z)$ successfully reproduces the corresponding term in \eqref{resultN2}.
There is another type of contribution summarized as
\begin{align}
\frac{jq}{c} f_a (-p,z) - \frac{j + q}{c}f_b (-p,z) + \frac{1}{c} f_b (-p,z) \, .
\end{align}
We read off the functions $f_a,f_b,f_c$ from \eqref{resultN2} as
\begin{align}
&f_a (-p,z) = 12 z^{2 j-p-1} \, _2F_1(-p,-p;-2 p;z) (\log (1-z)  (z-2)- 2 z ) \, , \nonumber \\
&f_b (-p,z) = - 12 p z^{2 j-p-1} ( z \, _2F_1(-p,-p;-2 p;z)- (z-1) \log (1-z) \, _2F_1(1-p,-p;-2 p;z)) \, , \nonumber \\
&f_c(-p,z) = 6 z^{2j -p-1} \bigl[ \, _2 F_1(-p,-p;-2 p;z) \left(\left(p^2 (z-1)+p (z-1)\right) \log (1-z)-p^2 z\right) \nonumber
\\ & \qquad \qquad  -p (z-1) \log (1-z) \, _2 F_1(1-p,-p;-2 p;z)\bigr]-z^{2j - p} r_p (z) \, .
\end{align}
The functions $f_a ,f_b$ reproduce those in \cite{Fitzpatrick:2016mtp}.
For the other function $f_c$, the terms proportional to $z^{2j - p + l}$ $(l=2,3,4,5)$ were obtained in \cite{Fitzpatrick:2016mtp}, and the $z$-expansion of the above result correctly reproduces them.
In other words, our method enables us to obtain the all-order expression in $z$ at the next non-trivial order in $1/c$, which was not available before.

One may wonder how this becomes possible. In \cite{Fitzpatrick:2016mtp}, the authors evaluated the operator product expansion (OPE) blocks in the sense of \cite{Czech:2016xec} in terms of open Wilson lines and then computed the four point Virasoro blocks using the expression of the OPE blocks.
The procedure requires the integration over $z_5, z_6$ for the two point function of the intermediate state as $\langle \mathcal{O}_p (z_5) \mathcal{O}_p (z_6) \rangle$, while
we need to integrate over only one parameter $x$ except for the positions $z_i$ of the currents inserted.
We think that this is a main reason for the simplicity.

\subsubsection{General W$_3$ blocks}
\label{largec}

So far we have analyzed general Virasoro blocks, but we also would like to make some comments on general W$_3$ blocks.
We are interested in the expectation value of $G_n (\Lambda_j|z_j)$ in \eqref{GnW}, where
we can obtain the vertex of open Wilson lines by solving the singlet condition in \eqref{singlet}.
Solutions to the condition were obtained in \cite{Fateev:2011qa} for $n=2,3,4$,
which may be denoted as
\begin{align}
\left[ \prod_{j=1}^n \langle X_j | \right] | S \rangle  = {\cal E}_n (\Lambda_j | X_j) 
\end{align}
for the $n$-point function.
For $n=3$, the representation $\Lambda = (\lambda_{1} , \lambda_{2})$ of an external operator is restricted to be degenerated as $\lambda_{1} =0$ or $\lambda_{2} =0$. A solution for $n=4$ is written as an integral of the product of two $n=3$ solutions, so there are also similar restrictions on the representations of external operators.

We first indicate that our formulation reduces to the one in \cite{Fateev:2011qa} at the leading order in $1/c$.
At this order, the expression in \eqref{GnW} reduces to
\begin{align}
\left. \langle G_3 (\Lambda_j | z_j) \rangle \right|_{\mathcal{O}(c^0)} = \left. \left[ \prod_{j=1}^3 e ^{z_{j}V_1^{2 }(X_j) } \right]
{\cal E}_n (\Lambda_j | X_j)  \right|_{X_j = 0} \, ,
\end{align}
where we have set $z_0 =0$ by using the independence of $z_0$.
Utilizing the formula \eqref{BCH}, we can see that the leading order expression is obtained from 
${\cal E}_n (\Lambda_j | X_j) $ by replacing $X_j = (x_j , y_j ,w_j)$ with $(z_j ,z_j , z_j^2/2)$.
Just like the $N=2$ case, the recipe to obtain the large $c$ limit of the W$_3$ block is the same as the one in \cite{Fateev:2011qa}.
As a result of this fact, we can conclude that our formulation with Wilson lines reproduces the large $c$ limit of W$_3$ blocks as obtained in \cite{Fateev:2011qa}, see also \cite{Poghosyan:2016lya}.
Similar analysis with Wilson lines was done in \cite{Besken:2016ooo} with a different technique for algebraic calculations. Our method with the $X$-basis makes computations easier for  generic representations as seen above, but their method seems to be convenient for a simple representation, such as, $\Lambda = \text{f}$. In appendix \ref{simple}, we compute the $1/c$ correction of a simple non-identity W$_3$ block from the Wilson line network \eqref{WLN} by utilizing their analysis.

We can also show that the analysis of two point function reduces to the previous one in subsection \ref{WilsonNew}. For $n=2$, the solution to the singlet condition \eqref{singlet} was obtained in \cite{Fateev:2011qa} as
\begin{align}
{\cal E}_2 (\Lambda , \bar{\Lambda} | X_1 , X_2) =2^{\lambda_1 + \lambda_2} \rho_{12}^{\lambda_1} \rho_{21}^{\lambda_2}   \, , \quad 
\rho_{ij} = y_i (x_i - x_j) - (w_i - w_j) \, .
\end{align}
For $z_0 =z_2$, the expression in \eqref{GnW} becomes
\begin{align}
G_2 (\Lambda , \bar{\Lambda} | z_1 , z_2) =
\left. W_{\Lambda} (z_1 ; z_2 ; X_1) \langle X_1 | \text{hw} \rangle  \right|_{X_1 = 0}
\end{align}
as for \eqref{Wh} with \eqref{fh}. Here, we have used
\begin{align}
\left. \rho_{12}^{\lambda_1} \rho_{21}^{\lambda_2} \right|_{X_2 = 0}
= \langle X_1 | \text{hw} \rangle \, , 
\end{align}
where $\langle X | \text{hw} \rangle$ was defined in \eqref{Xhw} with $\Lambda = (\lambda_1 , \lambda_2)$.
Similarly, for $z_0 = z_1$, we obtain 
\begin{align}
G_2 (\Lambda , \bar{\Lambda} | z_1 , z_2) =
\left. W_{\bar{\Lambda}} (z_2 ; z_1 ; X_2) \langle X_2 | \text{hw} \rangle\right|_{X_2 = 0} \, , 
\end{align}
but now $\langle X | \text{hw} \rangle$ is with  $\bar{\Lambda} = (\lambda_2 , \lambda_1)$.

\section{Heavy-light correlator}
\label{HLcorrelator}

The W$_N$ minimal model includes light operators of order $c^0$ and heavy operators of order $c$ as explained in section \ref{WNMM}.
Heavy operators correspond to non-trivial geometries,
and correlators with heavy operators are important for examining their quantum effects. 
In this section, we examine a heavy-light correlator for the simplest case with $N=2$,
and in section \ref{conclusion}, we make some comments on the general cases with $N \geq 3$.

For the analysis in this section, it is convenient to adopt the usual convention for the label of primary operators by two positive integers $(r,s)$, see \cite{DiFrancesco:1997nk} for a review.
In section \ref{WNMM}, we have used the numbers of boxes in Young diagrams as $(n_1^+ ; n_1^-)$ and the relation is $r = n_1^+ +1 , s =n_1^- + 1$. The conformal weight of primary operator $\mathcal{O}_{r,s}$ is given by \eqref{conformalweight} as
\begin{align}
\begin{aligned}
	h_{r,s} &= \frac{(r (k+3) - s (k+2) )^2 - 1}{4 (k+2) (k+3)}  \\
	&= - \frac{(s^2 - 1 ) c}{24} + \frac{1}{24} (- 12 rs + 13 s^2  - 1) + \mathcal{O} (c^{-1}) \, ,
	\end{aligned}
	\label{hrs}
\end{align}
which is of order $c$ for $n_1^- = s - 1 \neq 0$.
We study the four point correlator in \eqref{HLN} with $m=1$, $n = s -1$ for $N=2$, 
the expression with finite $k$ of which can be found in \eqref{HLN2}. 

It is also useful to work on the cylindrical coordinate $w$, which is related to the planar one as $z = \exp (i w)$. In particular, the conformal weight is shifted from $h_{r,s}$ in \eqref{hrs} to $h_{r,s} - \frac{c}{24}$.
The expression of the four point function in \eqref{HLN2} becomes%
\footnote{We have changed the phase factor for convenience, but it does not matter when we consider the product with the anti-holomorphic sector.}
\begin{align}
\begin{aligned}
	&\langle \mathcal{O}_{1,s} (- i \infty) \mathcal{O}_{2,1} (0) \mathcal{O}_{2,1} (w) \mathcal{O}_{1,s} (i \infty) \rangle  =   \frac{ (2 \sin \frac{s w}{2}) (- 2 i \sin \frac{w}{2})^{9/c }}{s}
\\
	&\qquad \qquad  - \frac{6 i }{c}
	e^{-\frac{ i s w}{2}} \sum_{l=1}^{s-1} \frac{(-1)^l (1-e^{i w})^{l+1} \Gamma (s)}{\Gamma (s- l) \Gamma (l+2)} \left( \sum_{j=1}^l \frac{1}{j} -  \frac{2 l}{l+1}  \right) + \mathcal{O} (c^{-2}) 
	\end{aligned}
	\label{gs2}
\end{align}
in the $1/c$ expansion.
In the rest of this section, we shall reproduce it from the bulk viewpoint.

\subsection{Conical spaces in AdS$_3$ gravity}

As explained in section \ref{WNMM}, $\mathcal{O}_{1,s}$ and $\mathcal{O}_{2,1}$ are supposed to be dual to a conical geometry and a bulk scalar field, respectively. For the heavy-light correlator in \eqref{gs2}, we evaluate the expectation value of the open Wilson line from $0$ to $w$ in the background of a conical defect sitting at the center of AdS$_3$.
In this subsection, we review the analysis on the quantum effects of conical spaces in \cite{Raeymaekers:2014kea}, which will be used in the next subsection for the bulk computation of the heavy-light correlator.

In the Chern-Simons formulation, a geometry is realized as a configuration of the gauge field.
We express the gauge field in the sl(2) Chern-Simons theory with the cylindrical coordinate as
\begin{align}
	a (w) = V_{1}^{2} - \frac{6}{c} J^{(2)} (w) V^{2}_{-1} 
\end{align}
instead of \eqref{az}.%
\footnote{The relative sign in front of $J^{(2)}$ (and the shift of conformal weight) can be explained from the transformation of the energy momentum tensor under the change of coordinates as \eqref{Tw} with \eqref{sd}.}
The conical defect at the center of AdS$_3$ corresponds to the classical configuration  as \cite{Raeymaekers:2014kea} $(s = 1,2,3,\ldots)$
\begin{align}
	L_0 = - \frac{ s^2 c }{24} \, , \quad L_{n \neq 0} = 0 
	\label{conical}
\end{align}
with the mode expansion of $J^{(2)}(w)$ as
\begin{align}
J^{(2)}(w) = \sum_{n} L_n e^{-i n w} \, .
\end{align}
In particular, the AdS vacuum is given by the case with $s = 1$.

Quantization of the conical spaces is analyzed in \cite{Raeymaekers:2014kea} utilizing the coadjoint orbits of the Virasoro algebra \cite{Witten:1987ty,Alekseev:1988ce}.
The conical defect at the center of AdS$_3$ is represented by operators inserted at $w = i \infty$ and $w = - i \infty$, and the corresponding states are denoted as $| 0 \rangle_s$ and ${}_s \langle 0 |$.
The states $| 0 \rangle_s$ and ${}_s \langle 0 |$ are defined as the Fock vacua of quantum operators $A_m$ as
\begin{align}
	A_m | 0 \rangle_s = 0 \, , \quad {}_s \langle 0 |A_{-m}  = 0 \, ,  \quad {}_s \langle 0 |0 \rangle_s = 1
\end{align}
with $m > 0$. Here $A_m$ satisfies
\begin{align}
	[A_m , A_n] = \frac{\delta_{m+n}}{m (m^2 - s^2)} \, .
\end{align}
The asymptotic symmetry is given by the Virasoro algebra generated by $L_n$, which satisfies
\begin{align}
	[L_m , L_n] = (m - n) L_{m+n} + \frac{c}{12} m^3 \delta_{m+n} \, .
	\label{Virasoroc}
\end{align}
These generators are written in terms of $A_m$ as
\begin{align}
	&L_0 = - \frac{ s^2 c }{24} + n_0 + \sum_{m =1}^\infty m^2 (m^2 - s^2) A_{-m} A_m + \mathcal{O} (c^{-1/2} )  \, ,
\end{align}
 see \cite{Raeymaekers:2014kea} for the others.
Requiring that  the generators satisfy \eqref{Virasoroc}, the number $n_0$ is determined as 
\begin{align}
	n_0 = \frac{1}{24} (s -1) (1 + 13 s) \, .
\end{align}
In particular, we have
\begin{align}
	L_m | 0 \rangle_s = 0 \, , \quad {}_s \langle 0 |L_{-m}  = 0 \, ,  \quad {}_s \langle 0 | L_0|0 \rangle_s = - \frac{s^2 c}{24} + n_0 + \mathcal{O} (c^{-1/2})
	\label{L0eig}
\end{align}
with $m > 0$. The $L_0$ eigenvalue is the same as $h_{1,s} - \frac{c}{24}$ in \eqref{hrs} up to the next non-trivial order in $1/c$, which confirms the  identification as $| 0 \rangle_s = \mathcal{O}_{1,s}(i \infty)|0 \rangle$ and ${}_s \langle 0 | = \langle 0 | \mathcal{O}_{1,s} ( - i \infty) $.

\subsection{Open Wilson line in conical space}

We consider the expectation value of the open Wilson line in the conical geometry with label $s$.
As explained in subsection \ref{Wilson}, we compute%
\begin{align}
	W_{s} (w)=\left . P \exp \left[ \int_0^w dw' \left( V_1^{2} - \frac{6}{c} J^{(2)}(w') V_{-1}^{2} \right) \right]  x \right|_{x=0} \, ,
\end{align}
where the background configuration for $J^{(2)} (w')$ is set as in \eqref{conical}.
The generators $V^{2}_n$ are given by \eqref{sl2gen} with $h_0 = -1/2$.
Here, we set the coupling in \eqref{ren} as $c_2=1$, which is enough for the analysis in this subsection.

We evaluate its expectation value by expanding in $1/c$.
Since $J^{(2)}(w')$ is now of order $c$, we define a new operator,
\begin{align}
\hat{J}^{(2)}(w') \equiv J^{(2)}(w') + \frac{s^2 c}{24} \, ,
\end{align}
which is of order $c^{1/2}$ as in \eqref{jj} below. 
As before, we find
\begin{align}
\begin{aligned}
	&\frac{d}{d w} \left[ e^{- (V^{2}_{1} + \frac{s^2}{4} V^{2}_{-1}) w} W_{s} (w) \right]
	\\& \quad = - \frac{6}{c} \hat{ J }^{(2)} (w)   (V^{2}_{-1})_s ( - w)
	\left[ e^{- (V^{2}_{1} + \frac{s^2}{4} V^{2}_{-1}) w} W_{s} (w) \right] 
\end{aligned}
\end{align}
with
\begin{align}
	(V_{-1}^2)_s ( w) \equiv e^{ (V_{1}^2 + \frac{s^2}{4} V_{-1}^2) w}  V_{-1}^2 e^{- (V_{1}^2 + \frac{s^2}{4} V_{-1}^2) w} \, .
\end{align}
Using the equation, the $1/c$ expansion is given by%
\begin{align}
	W_{h_0} (w)
	= \sum_{n=0}^\infty \left( - \frac{6}{c}\right)^n \int_0^w d w_n \cdots \int_0^{w_2} dw_1
	\left[ \prod_{j=1}^n \hat{ J} ^{(2)} (w_j)  \right] f_s (w_n , \ldots , w_1)
\end{align}
with
\begin{align}
	f_s (w_n , \ldots , w_1) =  \left .(V^2_{-1})_s (w_{*n}) \cdots  (V^2_{-1})_s (w_{*1}) e^{ (V^2_{1} + \frac{s^2}{4} V^2_{-1}) w}  x \right|_{x=0} \, , \quad w_{*i} \equiv w - w_i\, .
	\label{fs}
\end{align}
From (see (7.6) and (7.7) of \cite{Besken:2016ooo})
\begin{align}
	e^{(V^2_1 + \frac{s^2}{4} V^2_{-1}) w} = e^{c_1 (w) V^2_1} [c_0 (w)]^{2 V^2_0} e^{ c_{-1} (w) V^2_{-1}}
\end{align}
with
\begin{align}
	c_1 (w) = \frac{2}{s} \tan \frac{s w}{2} \, , \quad
	c_0 (w) = \cos \frac{s w}{2} \, , \quad
	c_{-1} (w) = \frac{s}{2} \tan \frac{s w}{2} \, ,
\end{align}
we find
\begin{align}
	e^{(V^2_{1} + \frac{s^2}{4} V^2_{-1}) w} x
	= c_0 (w) \left( x + c_1 (w) \right )
\end{align}
and
\begin{align}
	(V^2_{-1})_s ( w)  = c_0(w)^2 \left[ V^2_{-1} + 2 c_{1} (w)  V^2_0  + c_1 (w)^2 V^2_{1} \right] \, .
\end{align}
In particular, the leading order expression in $1/c$ is obtained as
\begin{align}
	\left. W_{s} (w) \right |_{\mathcal{O} (c^0)}= \frac{2}{s} \sin \frac{s w}{2} \, ,
	\label{tree}
\end{align}
which reproduces \eqref{gs2}.

 At the next leading order in $1/c$, there are two types of contributions with one and two insertions of $\hat {J}^{(2)}$.
With one insertion, we compute
\begin{align}
\begin{aligned}
 H_s^{(1)} (w) &= -	\frac{6}{c} \int_0^w  dw_1    f_s (w_1) {}_s \langle 0 |\hat{ J }^{(2)}(w_1) | 0 \rangle_s \\
& = \frac{12 n_0}{c } \left[ \frac{2\sin \left(\frac{s w}{2}\right) - s w \cos \left(\frac{s w}{2}\right) }{s^3} \right] \, ,
\end{aligned}
\label{HL1}
\end{align}
where we have used
\begin{align}
\begin{aligned}
	&{}_s \langle 0 | \hat{J}^{(2)}(w_1) | 0 \rangle_s =  n_0  + \mathcal{O} (c^{-1}) \, ,
\end{aligned}
\end{align}
see \eqref{L0eig}.
Here we do not need to worry about the regularization issue.
In the following, we examine the case of two insertions with special care on the issue.

With the two insertions of $\hat{J}^{(2)}$, the integral is
\begin{align}
	\left(-\frac{6}{c} \right)^2 \int_0^w  dw_2  \int_0^{w_2}  dw_1 f_s (w_2,w_1) \,	 {}_s \langle 0 | \hat {J }^{(2)} (w_2)  \hat {J}^{(2)}(w_1) | 0 \rangle_s 
   \label{twoTp}
\end{align}
with 
\begin{align}
f_s (w_2,w_1) =  \frac{8}{s^3} \sin \left(\frac{s w_1}{2}\right) \sin \left(\frac{s (w- w_2)}{2} \right) \sin \left(\frac{  s (w_2 -w_1)}{2}\right) \, .
\label{fs2}
\end{align}
Moreover, we find
\begin{align}
\begin{aligned}
&\left.  {}_s \langle 0 | \hat{J}^{(2)}(w_1) \hat{J}^{(2)} (w_2) | 0 \rangle_s \right|_{\mathcal{O}(c)}=  \sum_{m,n > 0} e^{- i(m w_1 - n w_2)} \left.
{}_s \langle 0 | [ L_m  , L_{-n} ]| 0 \rangle_s \right|_{\mathcal{O}(c)}  \\
& \quad = \frac{c}{12} \sum_{m > 0} e^{- i m (w_1 - w_2)} (m^3 - s^2 m)
= \frac{c}{16} \left( \frac{1}{2 \sin ^4  \left( \frac{w_1 - w_2}{2}\right)} +  \frac{s^2 -1}{3 \sin ^2 \left( \frac{w_1 - w_2}{2}\right)} \right) \, .
\end{aligned}
\label{jj}
\end{align}
However, the integration over $w_1,w_2$ diverges in \eqref{twoTp}, and we have to properly regularize it.

In order to make the expression simpler, we introduce new parameters as
\begin{align}
	z = e^{ i w} \, , \quad z_1 = e^{i w_1} \, , \quad z_2 = e^{i w_2} \, , \quad
	y = e^{ i s w} \, , \quad y_1 = e^{i s w_1} \, , \quad y_2 = e^{i s w_2} \, .
	\label{zy}
\end{align}
Then, the expression in \eqref{jj} can be rewritten as
\begin{align}
\begin{aligned}
	&\left. {}_s \langle 0 | \hat{J}^{(2)} (w_1) \hat{J}^{(2)} (w_2) | 0 \rangle_s \right|_{\mathcal{O}(c)}=
\frac{c}{2} \frac{z_1^2 z_2^2}{(z_1 - z_2)^4} - \frac{c}{12} \frac{(s^2 -1) z_1 z_2}{(z_1 -z_2)^2} \, .
\end{aligned}
\label{TT}
\end{align}
The integral \eqref{twoTp} diverges due to $(y_1 - y_2)^{-4}$ singularity in \eqref{TT} as discussed in \cite{Fitzpatrick:2015dlt}, so we separate the singular part in \eqref{TT} as
\begin{align}
\begin{aligned}
	&\left. {}_s \langle 0 | \hat{J}^{(2)} (w_1) \hat{J}^{(2)}(w_2) | 0 \rangle_s \right|_{\mathcal{O}(c)} =
	g_1[w_1 , w_2] + g_2 [w_1 ,w_2] \, , \\
	&\frac{1}{s^4 (z_1 z_2)^{2s}} g_1 [w_1 , w_2] =  \frac{c (z_1 z_2 )^{2-2 s}}{2 s^4 (z_2 - z_1)^4}
	-\frac{c \left(s^2-1\right) (z_1 z_2)^{1-2 s}}{12 s^4 (z_2 - z_1)^2} -\frac{c }{2 \left(y_1 - y_2\right)^4} \, , \\
	&
	\frac{1}{s^4 (z_1 z_2)^{2s}} g_2 [w_1 , w_2] =
	\frac{c }{2 \left(y_1 - y_2\right)^4 } \, .
\end{aligned}
\end{align}
The integral with $g_1[w_1,w_2]$ was performed in \cite{Fitzpatrick:2015dlt}, which yields a finite result.
The integral with $g_2[w_1 ,w_2]$ diverges so we need to regularize it in a proper way.

The integral with $g_1[w_1,w_2]$ can be written as
\begin{align}
\begin{aligned}
 H^{(2)}_s (w)
   =&  \frac{6^2 i z^{-s/2}}{c s^3} \int_1^z d z_2 \int_1^{z_2} d z_1 z_1^{s-1}z_2^{s-1} \left(1-z_1^s\right) \left(z_2^s-z^s\right) \left(z_1^s-z_2^s\right) \\
	&\times  \left(\frac{(z_1 z_2)^{2-2 s}}{2 (z_2-z_1)^4} -\frac{\left(s^2-1\right) (z_1 z_2)^{1-2 s}}{12 (z_2-z_1)^2} -\frac{s^4}{2 \left(z_1^s-z_2^s\right)^4} \right) 
\end{aligned}
\label{HL2}
\end{align}
by using the explicit form of $f_s(w_2,w_1)$ in \eqref{fs2}.
The expression with general complex $s$ after the integration over $z_1,z_2$ was obtained in \cite{Fitzpatrick:2015dlt}. However, several terms diverge for positive integer $s$, and we should take care of cancellations among them.
Instead of doing so, we evaluate the integral directly with the explicit values of $s$.
As for illustration, we write down the expressions for $s=2,3,4$ as
\begin{align}
	&	\frac{ c s^3}{6^2 i z^{-s/2}} H_2^{(2)} (w) = \frac{1}{8} \left(\left(1-7 z^2\right) \log (z)-\left(z^2-1\right) (-8 \log (z+1)-3+ 8 \log (2))\right) \, , \nonumber \\
	&\frac{ c s^3}{6^2i  z^{-s/2}} H_3^{(2)} (w) = \frac{1}{12} \left(\left(7-47 z^3\right) \log (z)+z \left(z^2 \left(\frac{49}{3}-27 \log (3)\right)-9 z+9\right) \right. \ \nonumber \\ & \qquad \left. +27 \left(z^3-1\right) \log \left(z^2+z+1\right)-\frac{49}{3}+27 \log (3)\right) \, ,  \\
	& \frac{ c s^3}{6^2 i z^{-s/2}} H_4^{(2)} (w) = \frac{1}{12} \left(-96 \left(z^4-1\right) \log (2)+\left(19-125 z^4\right) \log (z)+48 \left(z^4-1\right) \log (z+1) \right.  \nonumber \\ & \qquad \left. +48 \left(z^4-1\right) \log \left(z^2+1\right)+\frac{1}{6} (z-1) (z+1) (z (223 z-128)+223)\right) \, . \nonumber 
\end{align}
In order to compare with \eqref{gs2}, we should set $z= \exp ( i w)$ at the end of computation.

A main problem here is to regularize the divergence in the integral with  $g_2[w_1 ,w_2]$, which can be written as
\begin{align}
	& \frac{18 i y^{-1/2}}{c s} \int_1^y dy_2 \int_1^{y_2} dy_1
	\frac{(1 - y_1) (y_2 - y) (y_1 - y_2)}{(y_2 - y_1)^{4 }}
\end{align}
in terms of $y,y_1,y_2$ in \eqref{zy}.
Notice that this is proportional to the $1/c$ correction in
$
\left. \langle W_{-1/2} (y , 1) \rangle \right|_{\mathcal{O}(c^{-1})}
$ with the AdS$_3$ background. Applying the regularization for the case, we find
\begin{align}
\begin{aligned}
	&\frac{18 i y^{-1/2}}{c s} \int_1^y dy_2 \int_1^{y_2} dy_1
	\frac{(1 - y_1) (y_2 - y) (y_1 - y_2)}{(y_2 - y_1)^{4 - 2 \epsilon}}
	 \\ & \qquad
	 =\frac{18 i y^{-1/2}}{c s} \left[ \frac{1-y}{4 \epsilon}-\frac{1}{4} (y-1) (2 \log (y-1)+1)\right] + \mathcal{O} (\epsilon) \, .
\end{aligned}
\end{align}
However, this computation is too naive since we are actually working on the cylindrical coordinate $w$ but not on  $y (\equiv \exp ( i sw))$. Using the relation (see appendix \ref{Shwarz} for some discussions)
\begin{align}
	\left. \langle W_{-1/2} (w , 0) \rangle \right|_{\mathcal{O}(c^{-1})}
	= \left. \left[ \left( \frac{\partial w}{\partial y} \right)^{-h_{1,2}}
	 \langle W_{-1/2} (y , 1) \rangle \right]  \right|_{\mathcal{O}(c^{-1})} \, ,
\end{align}
we find
\begin{align}
\begin{aligned}
	& \frac{18 i y^{-1/2}}{c s} \int_1^y dy_2 \int_1^{y_2} dy_1
	\frac{(1 - y_1) (y_2 - y) (y_1 - y_2)}{(y_2 - y_1)^{4 }}
 \\ & \qquad
	 \sim - \frac{9 i y^{-1/2}}{ c s} (y-1) ( \log (y-1) - \tfrac12 \log (y)  +  \delta )  \equiv H_s^{(3)} (w)\, .
\end{aligned}
\label{HL3}
\end{align}
Here, we have  removed the divergent term by changing the overall normalization of the open Wilson line. Moreover, we introduce $\delta$ by using the ambiguity at the $\epsilon^0$ order.
For the comparison with \eqref{gs2}, we should set $y = \exp (i s w)$ as above.

The $1/c$ order correction to the tree level result in \eqref{tree} is given by the sum of three contributions as $\sum_{a=1}^3 H_s^{(a)} (w)$, which are given in \eqref{HL1}, \eqref{HL2}, and \eqref{HL3}.
We have checked for explicit values of $s$ (i.e., for $s=2,3,\ldots,50$)
that the sum reproduces the $1/c$ order expression of \eqref{gs2} by setting the parameter as $\delta = - \log (s)$.

\section{Conclusion and open problems}
\label{conclusion}

We evaluated the networks of open Wilson lines in the $\text{sl}(N) $ Chern-Simons gauge theory with $N=2,3$ including quantum corrections. We compared the results with the $1/c$ expansions of correlators and conformal blocks in the W$_N$ minimal model \eqref{coset}. In order to treat divergences associated with loop diagrams, we adopted the renormalization prescription proposed in \cite{Hikida:2017ehf}.
In the previous paper, we examined the two and three point functions in \eqref{23pt} with general light operators for $N=2$ and with specific operators for $N=3$ by applying the prescription. 
We first extended the analysis by evaluating the correlators in \eqref{23pt} with general light operators for $N=3$. Applying the method, the identity W$_3$ block was computed from the product of two open Wilson lines, and the result reproduced the CFT one obtained in appendix \ref{toda}. We further studied general conformal blocks with $N=2$. In particular, we obtained the all-order expressions in $z$ expansion at the $1/c$ order, which are available in \cite{Fitzpatrick:2016mtp} except for a part. 
We also made some comments on the $N=3$ extensions.
For a heavy-light correlator, we examined an open Wilson line in a conical geometry, where we utilized the analysis in \cite{Raeymaekers:2014kea} for the quantum effects of the geometry. The analysis may be understood as a bulk interpretation of the CFT computations in \cite{Fitzpatrick:2015dlt}. 
We concentrated on the Virasoro case for the $1/c$ corrections of general light-light blocks and heavy-light correlators, but it is desired to analyze the W$_3$ case in a systematic way.

For the general W$_3$ blocks at the leading order in $1/c$, we explained that our formulation nicely combines previous analyses in \cite{Fateev:2011qa} from CFT and in \cite{Besken:2016ooo} with open Wilson lines.
We thought that it is straightforward to extend the analysis beyond the leading order just like the $N=2$ examples, but it does not seem to be the case.
A complication may arise from the use of the $X$-basis as, for instance, the bra and ket states are expressed in a quite different way as in \eqref{Xlw} and \eqref{Xhw}.
In appendix \ref{simple}, we analyze the $1/c$ correction of a simple example using the technique of \cite{Besken:2016ooo}, but we would like to develop our method with the $X$-basis for more systematic analysis. 

The analysis of a heavy-light correlator is also much simpler with $N=2$.
For general $N$, we may use the results in \cite{Campoleoni:2017xyl}
on the quantum effects of conical spaces.  However, it is non-trivial to regularize divergences from loops associated with the open Wilson line in a non-trivial background. 
For $N=2$, a four point function in the heavy-light limit can be mapped to a two point function by applying a coordinate transformation as explained in \cite{Fitzpatrick:2015zha,Fitzpatrick:2015dlt}. Thus, we can apply our regularization prescription in the new coordinate system and come back to the original setup, see appendix \ref{Shwarz}. However, we do not know whether we have such a nice transformation for general $N$. 

The current analysis can be applied to other setups.
As a next project, we would like to consider supersymmetric extensions.
These extensions are important to connect higher spin theory to superstring theory as in \cite{Creutzig:2013tja,Creutzig:2014ula,Hikida:2015nfa} 
with $\mathcal{N}=3$ supersymmetry and \cite{Gaberdiel:2013vva,Gaberdiel:2014cha} for $\mathcal{N}=4$ supersymmetry.
As an example, the $\mathcal{N}=2$ version of \cite{Gaberdiel:2010pz} was proposed in \cite{Creutzig:2011fe}, and the large $c$ regime of the $\mathcal{N}=2$ W$_{N+1}$ minimal model is claimed to be described by the Chern-Simons theory based on $\text{sl}(N+1|N)$ \cite{Hikida:2012eu}.
Therefore, superconformal blocks are expected to be computed by the networks of open Wilson lines in a Chern-Simons theory based on a superalgebra.

An important point of our formulation is to use open Wilson lines instead of bulk scalar propagators.
It is better to use bulk scalars if we want to do higher dimensional extensions, see, e.g.,  \cite{Maxfield:2017rkn} for a recent argument. However, Wilson lines are gauge invariant objects and should be useful in examining the properties of geometry in terms of Chern-Simons theory. In particular, a Wilson line was used to compute the entanglement entropy of boundary theory in a holographic way \cite{deBoer:2013vca,Ammon:2013hba}, which is a higher spin extension of \cite{Ryu:2006bv,Ryu:2006ef}, see also \cite{deBoer:2014sna}.
We expect that our formulation is useful to extract non-trivial information on the quantum effects of non-trivial geometry.

\subsection*{Acknowledgements}

We are grateful to Changhyun Ahn, Andrea Campoleoni, and Euihun Joung for useful discussions.
We thank the organizers of ``East Asia Joint Workshop on Fields and Strings 2017, KEK Theory Workshop 2017,''  and Y. H. thanks the organizers of the workshop ``New Ideas on
Higher Spin Gravity and Holography'' at Kyung Hee University, Seoul for their hospitality.
The work of Y. H. is supported by JSPS KAKENHI Grant Number 16H02182.

\appendix
\section{CFT calculations}
\label{toda}

In this appendix, we first provide the computation of the four point function \eqref{4pt0} with $\Lambda = n \bm{\omega}_m$ and $\Xi = \text{f}$, i.e.,
\begin{align}
G(z) \equiv \langle \bar{\cO}_{(n \bm{\omega}_m;0)}(\infty) \cO_{(n \bm{\omega}_m;0)}(1) \bar{\cO}_{(\text{f};0)}(z) \cO_{(\text{f};0)}(0) \rangle \, .
\label{fourpf}
\end{align}
We adopt the Coulomb gas method in \cite{Papadodimas:2011pf,Hijano:2013fja,DiFrancesco:1997nk}, see also \cite{Fateev:2007ab,Fateev:2008bm,Chang:2011vka}.
We then compute the three point functions in \eqref{23pt} as
\begin{align}
\langle \bar{\cO}_{(n \bm{\omega}_m;0)}(z_1) \cO_{(n \bm{\omega}_m;0)}(z_2) J^{(s)}(z_3) \rangle 
\label{threepf}
\end{align}
from the conformal block decomposition of the four point function \eqref{fourpf} by following the strategy in \cite{Hikida:2017byl,Hikida:2017ehf}.

%%%%%%%%%%%%%%%%%%%%%%%%%%%%%%%%%%%%%%%%%%%%%%%%%%%%%%%%%%%%%%%%%%%%%%%%%%
\subsection{Coulomb gas method}

We start from $N-1$ free scalar fields $\bm{\phi}\equiv\phi^i \; (i=1,\dots,N-1)$ with the normalization of two point functions as
\begin{align}
\langle \phi^i(z,\bar{z}) \phi^j(0,0) \rangle = -2\delta^{ij}\log|z|^2 \; .
\end{align}
We use the vertex operators
\begin{align}
V_{\bb} = : e^{i\bb\cdot\bm{\phi}} : \, , \quad 
\bb = -\alpha_+\Lambda_+ -\alpha_-\Lambda_- 
\label{brep}
\end{align}
with the conformal weight $h_{\bb}=\bb^2$ and $\alpha_\pm$ in \eqref{alphapm}.
The su$(N)$ weight vectors $\Lambda_\pm$ are decomposed as
\begin{align}
\Lambda_\pm = \sum_{i=1}^{N-1}\lambda_i^\pm\bm{\omega}_i
\end{align}
with the Dynkin labels $\lambda_ i ^\pm$ and the fundamental weights $\bm{\omega}_i$. 
The Dynkin labels are related to the numbers $n_i^\pm$ introduced in section \ref{WNMM} as 
$\lambda^\pm_i = n^\pm_i - n_{i+1}^\pm$.
In particular, we set
\begin{align}
\begin{aligned}
\bb_{(\text{f};0)} = -\alpha_+\bm{\omega}_1 \;,\quad
\bb_{(n\bm{\omega}_m ; 0)} = -\alpha_+n\bm{\omega}_m 
\label{map}
\end{aligned}
\end{align}
for our computations.

We introduce a background charge, which modifies the energy momentum tensor as
\begin{align}
T(z) = -\frac{1}{4} \partial\bm{\phi} \cdot \partial\bm{\phi} + i \bm{\alpha}_0 \cdot \partial^2 \bm{\phi} \, .
\end{align}
Here $\bm{\alpha}_0 \equiv \alpha_0 \bm{\rho}$ with $\alpha_0 = \alpha_+ + \alpha_-$ and $\bm{\rho} \, (=\sum_i \bm{\omega}_i)$ is the Weyl vector of su$(N)$.
Then, the conformal weight $h_{\bb}$ is shifted as
\begin{align}
h_{\bb}=\bb\cdot(\bb-2\bm{\alpha}_0) \, .
\end{align}
This is invariant under the exchange $\bb \leftrightarrow 2\bm{\alpha}_0-\bar{\bb}$,%
\footnote{The conjugation operation $\bar{\bb}$ replaces $\lambda^\pm_i$ by $\lambda^\pm_{N-i}$.}
which implies that we should identify the operator $V_{\bb}$ with $V_{2\bm{\alpha}_0- \bar{\bb}}$. 
In terms of weight vectors, the conformal weight becomes \eqref{conformalweight}, and in particular,
\begin{align}
\begin{aligned}
&h(\text{f};0)  = \frac{N-1}{2N} \left( 1+\frac{N+1}{k+N} \right) \;, \quad
&h(n \bm{\omega}_m;0) = \frac{nm(N-m)}{2N} \left( n + \frac{N+n}{k+N} \right)
\end{aligned}
\end{align}
with finite $k,N$.

In the free theory, the correlator of the form
\begin{align}
\langle V_{\bb_1}(z_1) \dots V_{\bb_n}(z_n) \rangle
\end{align}
is non-zero when the charges are conserved.
In the presence of background charge, the conservation  rule is  shifted as
\begin{align}
\sum_{i=1}^n \bb_i = 2 \bm{\alpha}_0 \;.
\label{chargeconserve}
\end{align}
In order to satisfy this condition and compute a correlation function, we introduce the screening operator
\begin{align}
Q \equiv \oint dz V_{\tilde{\bb}}(z) \, ,
\end{align}
where $\tilde{\bb}$ is chosen to satisfy $h_{\tilde{\bb}} = 1$.
The condition comes from the fact that the dimension of $Q$ should be zero in order not to break conformal properties of the correlation function. 
From this condition, we find
\begin{align}
\tilde{\bb} = \alpha_\pm \bm{e}_i \, ,
\end{align}
where $\bm{e}_i$ is one of the simple roots of su$(N)$.
For our calculations, it is useful to use the following relations about the inner products as
\begin{align}
\begin{aligned}
&\bm{\omega}_i \cdot \bm{\omega}_{j} = \frac{(N-i)j}{N}~~(i \geq j) \;,\quad
\bm{\omega}_i \cdot \bm{e}_j = \delta_{ij} \;,\\
&\bm{e}_i \cdot \bm{\rho} = 1 \;,\quad
\bm{e}_i \cdot \bm{e}_j = A_{ij} \;,
\label{relation}
\end{aligned}
\end{align}
where $A_{ij}$ is the Cartan matrix of su$(N)$.

%%%%%%%%%%%%%%%%%%%%%%%%%%%%%%%%%%%%%%%%%%%%%%%%%%%%%%%%%%%%%%%%%%%%%%%%%%
\subsection{Four point function}

We consider the four point function in \eqref{fourpf}. From the mapping \eqref{map}, the correlator is written as
\begin{align}
G(z) = \left\langle V_{2\bm{\alpha}_0+\alpha_+ n\bm{\omega}_{m}}(z_\infty) V_{-\alpha_+ n\bm{\omega}_{m}}(1) V_{-\alpha_+ \bm{\omega}_{N-1}}(z) V_{-\alpha_+ \bm{\omega}_{1}}(0) \left(\prod_{i=1}^{N-1}\oint dt_i V_{\alpha_+\bm{e}_i}(t_i)\right) \right\rangle \, .
\end{align}
Here we have introduced $z_\infty$, which will be set as $z_\infty = \infty$ at the end of the computation. 
We have changed the first operator $V_{\bb} \rightarrow V_{2\bm{\alpha}_0-\bar{\bb}}$ and
inserted the screening operators to satisfy (\ref{chargeconserve}).
Using the simple form of the free field correlator as 
\begin{align}
\langle V_{\bb_1}(z_1) \dots V_{\bb_n}(z_n) \rangle = \prod_{i<j} |z_i-z_j|^{4\bb_i\cdot\bb_j} \;,
\end{align}
the holomorphic part of $G(z)$ is written as
\begin{align}
& G(z) =(z_\infty-1)^{\gamma_1} (z_\infty-z)^{\gamma_z} z_\infty^{\gamma_0} (1-z)^{2\alpha_+^2\frac{nm}{N}} z^{2\alpha_+^2\frac{1}{N}} \nonumber \\
& \quad \times \oint dt_1 \cdots \oint dt_{m-1} t_1^{-2\alpha_+^2} \left[ \prod_{i=1}^{m-1} (t_i-z_\infty)^{4\alpha_0\alpha_+} (t_i-t_{i+1})^{-2\alpha_+^2} \right] \nonumber \\
&\quad \times \oint dt_{m} (t_{m}-z_\infty)^{4\alpha_0\alpha_+ + 2n\alpha_+^2} (t_{m}-1)^{-2n\alpha_+^2}\\
&\quad \times \oint dt_{m+1} \cdots \oint dt_{N-1}( t_{N-1} - z)^{-2\alpha_+^2} \left[ \prod_{i=m+1}^{N-1} (t_i-z_\infty)^{4\alpha_0\alpha_+} \right] \left[ \prod_{i=m}^{N-2} (t_i-t_{i+1})^{-2\alpha_+^2} \right] \; . \nonumber
\end{align}
Here, we can compute $\gamma_1,\gamma_x,\gamma_0$ explicitly, but they are not necessary for our purpose. Using the relation
\begin{align}
\oint dt_i \prod_{j=1}^{3} (t_i-s_j)^{2\beta_j} \propto (s_1-s_2)^{\beta_1+\beta_2-\beta_3}(s_2-s_3)^{\beta_2+\beta_3-\beta_1}(s_3-s_1)^{\beta_3+\beta_1-\beta_2}
\end{align}
for $\sum_{j=1}^{3} \beta_j = -2$, we can integrate over all variables without $t_{m}$.
After the integration, we arrive at
\begin{align}
G(z) \propto \left| z^{2\alpha_+^2\frac{1}{N}} (1-z)^{2\alpha_+^2\frac{nm}{N}} \oint dt_{m} t_{m}^a (t_{m}-1)^b (t_{m}-z)^c \right|^2
\end{align}
with
\begin{align}
a = m-1-2m\alpha_+^2 \, , \quad
b = -2n\alpha_+^2 \, , \quad c= N-m-1-2(N-m)\alpha_+^2 \;.
\end{align}
As in (9.88) of \cite{DiFrancesco:1997nk}, the final result should be of the form
\begin{align}
\begin{aligned}
& \left| \oint dy y^a (y-1)^b (y-z)^c \right|^2  \\
 & \qquad = \frac{\sin(\pi b)\sin(\pi(a+b+c))}{\sin(\pi(a+c))} |I_1(z)|^2 + \frac{\sin(\pi a)\sin(\pi c)}{\sin(\pi(a+c))} |I_2(z)|^2 \, ,
 \end{aligned}
\end{align}
where $I_i(z)$ are
\begin{align}
\begin{aligned}
I_1(z) &= \frac{\Gamma(-a-b-c-1) \Gamma(b+1)}{\Gamma(-a-c)} {}_2F_1(-c, -a-b-c-1; -a-c; z) \;,\\
I_2(z) &= z^{1+a+c} \frac{\Gamma(a+1) \Gamma(c+1)}{\Gamma(a+c+2)} {}_2F_1(-b, a+1; a+c+2; z) \;.
\end{aligned}
\end{align}
Applying this result and using the relation
\begin{align}
{}_2F_1(a, b; c; z) = (1-z)^{c-a-b} {}_2F_1(c-a, c-b; c; z) \;,
\label{Kummer}
\end{align}
we obtain \eqref{coulomb} with \eqref{CoulombN}.
%%

%%%%%%%%%%%%%%%%%%%%%%%%%%%%%%%%%%%%%%%%%%%%%%%%%%%%%%%%%%%%%%%%%%%%%%%%%%%%%%%%%%%%%%%%%%%%%%%%%%%%%%%%

\subsection{Identity block expansion and higher spin charges}

We consider the holomorphic part of the identity block as
\begin{align}
 z^{-2 h(\text{f};0) } (1-z)^{nm\frac{k+N+1}{N(k+N)}} {}_2F_1\left(n+\frac{n}{k+N},-\frac{m}{k+N};-\frac{N}{k+N};z \right) \;.
\label{vb}
\end{align}
We obtain the three point functions in \eqref{threepf} including 1/$c$ corrections by applying the method in \cite{Hikida:2017byl,Hikida:2017ehf}. We first decompose the four point function as
\begin{align}
|z|^{4 h(\text{f};0) } G(z) = \mathcal{V}_0(z) + \sum_{s=3}^\infty C_\text{f}^{(s)} C_{n \bm{\omega}_m} ^{(s)}\mathcal{V}_s(z) + \dots \, .
\label{dec}
\end{align}
Here $\mathcal{V}_0(z)$  and $\mathcal{V}_s(z)$ are the identity Virasoro block and the Virasoro block of the spin $s$ current, and their expansions in $1/c$ are given in \eqref{V0} and \eqref{Vp} with $h_p = s$. The coefficients $C^{(s)}_\Lambda$ are defined as
\begin{align}
C^{(s)} _\Lambda \equiv \frac{\langle \bar{\mathcal{O}}_{(\Lambda ; 0)} \mathcal{O}_{(\Lambda ; 0)}  J^{(s)} \rangle}{\langle J^{(s)} J^{(s)} \rangle^{1/2}   \langle \bar{\mathcal{O}}_{(\Lambda ; 0)} \mathcal{O}_{(\Lambda ; 0)} \rangle   } \; , \label{Cs}
\end{align}
which is expanded in $1/c$ as
\begin{align}
C^{(s)}_\Lambda = c^{-1/2} \left[ C_{\Lambda, 0}^{(s)} +c^{-1}C_{\Lambda,1}^{(s)} +\mathcal{O}(c^{-2}) \right] \;. \label{Csexp}
\end{align}
On the other hand, from  \eqref{kexp},
the expansion of \eqref{vb} in 1/$c$ is obtained as
\begin{align}
\begin{aligned}
&|z|^{4 h(\text{f};0) } G(z)  \\
&\sim 1+ \frac{(1-N^2)}{c} \sum_{l=1}^\infty \left( -\frac{nm}{l} + \frac{n}{\Gamma(m)} \frac{\Gamma(1+N)\Gamma(l+m)}{l\Gamma(l+N)} \right) z^l + \frac{1}{c^2} \sum_{l=2}^\infty f_{c;n\bm{\omega}_m}^{(l)} z^l \, ,
\end{aligned}
\label{1/cexp0}
\end{align}
where we have defined
\begin{align}
&\frac{f_{c;n\bm{\omega}_m}^{(l)}}{(1-N^2)^2}  \nonumber \\&= \frac{nN^2\Gamma(N)\Gamma(l+m)}{l\Gamma(m)\Gamma(l+N)} \left[ \sum_{j=1}^{l} \frac{N}{N+j-1} - \left( m \sum_{j=1}^l \frac{1}{j+m-1} -n H_{l-1} \right) - \frac{2N^2+2N+1}{N(N+1)} \right] \nonumber \\
&- n^2m \frac{N\Gamma(N)}{\Gamma(m)} \sum_{j=1}^{l-1} \frac{1}{l-j} \frac{\Gamma(j+m)}{j\Gamma(j+N)} + \frac{nm}{l} \left( 2N + \frac{1}{N+1} +nmH_{l-1} \right) \;.
  \label{1/cexp} 
\end{align}
Comparing the $1/c$ and $z$ expansions in the both sides of \eqref{dec}, we obtain constraints for the coefficients $C^{(s)}_\Lambda$.

We find the explicit form of the coefficients $C^{(s)}_\Lambda$ by solving the constraint equations.
From the coefficients of $z^l$, constraints at the leading order in $1/c$ are given as
\begin{align}
\left( -\frac{nm}{l} + \frac{n}{\Gamma(m)} \frac{\Gamma(1+N)\Gamma(l+m)}{l\Gamma(l+N)} \right) = \frac{ (\Gamma(l))^2}{1-N^2} \sum_{s=2}^l \frac{\Gamma(2s) C_{\text{f},0}^{(s)} C_{n\bm{\omega}_ m,0}^{(s)}}{(\Gamma(s))^2 \Gamma(s+l) (l-s)!} \, ,
\end{align}
where
\begin{align}
C_{\text{f},0}^{(2)} C_{n\bm{\omega}_ m,0}^{(2)}= \frac{nm}{2} (1-N)(m-N)
\end{align}
is read off from \eqref{hexpN}.
The first few solutions to the constraint equations are
\begin{align}
\begin{aligned}
&C_{\text{f},0}^{(3)} C_{n\bm{\omega}_ m,0}^{(3)}= \frac{nm(1-N)(m-N)(2m-N)}{6(N+2)} \;,\\
&C_{\text{f},0}^{(4)} C_{n\bm{\omega}_ m,0}^{(4)} =\frac{nm (1-N) (m-N) \left(5 m^2-5 m N+N^2+1\right)}{20 (N+2) (N+3)} \;,\\
&C_{\text{f},0}^{(5)} C_{n\bm{\omega}_ m,0}^{(5)} =\frac{nm(1-N) (m-N) (2 m-N) \left(7 m^2-7 m N+N^2+5\right)}{70 (N+2) (N+3) (N+4)} \;.
\end{aligned}
\end{align}
Using the definition of the coefficients $C_{\Lambda,0}^{(s)}$ in \eqref{Csexp} with \eqref{Cs} and the normalization of currents in \eqref{2ptcurrent} with \eqref{BsNs}, we rewrite the three point functions as
\begin{align}
\left. w^{(s)} (\Lambda , 0) \right|_{\mathcal{O}(c^0)} \equiv
\left. \langle \bar{\mathcal{O}}_{(\Lambda ; 0)} \mathcal{O}_{(\Lambda ; 0)} J^{(s)}  \rangle \right|_{\mathcal{O}(c^0)} =
\frac{\Gamma (N+s)}{(1-N) \Gamma (N+2)} C_{\text{f},0}^{(s)} C_{\Lambda,0}^{(s)} \, ,
\end{align}
which leads to
\begin{align}
\begin{aligned}
&\left. w^{(3)} (n\bm{\omega}_ m , 0) \right|_{\mathcal{O}(c^0)}= \frac{1}{6} m n (m - N) (2 m - N) \, , \\
&\left. w^{(4)} (n\bm{\omega}_ m , 0) \right|_{\mathcal{O}(c^0)} = \frac{1}{20} m n (m-N) \left(5 m^2-5 m N+N^2+1\right) \, ,\\
&\left. w^{(5)} (n\bm{\omega}_ m , 0) \right|_{\mathcal{O}(c^0)} = \frac{1}{70} m n (m-N) (2 m-N) \left(7 m^2-7 m N+N^2+5\right) \, .
\end{aligned}
\end{align}
We can see that the results with $s=3,4$ and general $n,m$ are the same as \eqref{wexpN} and (C.3) in \cite{Gaberdiel:2011zw}.

At the next order in $1/c$, constraint equations for $s=3,4,5$ are given as
\begin{align}
&f_{c;n\bm{\omega}_ m}^{(3)} = f_{c;n\bm{\omega}_ m}^{(2)} + \left( C_{\text{f},0}^{(3)}C_{n\bm{\omega}_ m,1}^{(3)} + C_{\text{f},1}^{(3)}C_{n\bm{\omega}_ m,0}^{(3)} \right)\, , \nonumber \\
&f_{c;n\bm{\omega}_ m}^{(4)} = f_{c;n\bm{\omega}_ m}^{(2)}\frac{9}{10} + \frac{(1-N)^2}{8(1+N)^2} + \frac{1-N}{10(1+N)^2} + \frac{1}{50(1+N)^2} \nonumber \\
&  + \left( C_{\text{f},0}^{(3)}C_{n\bm{\omega}_ m,1}^{(3)} + C_{\text{f},1}^{(3)}C_{n\bm{\omega}_ m,0}^{(3)} \right)\frac{3}{2} + \left( C_{\text{f},0}^{(4)}C_{n\bm{\omega}_ m,1}^{(4)} + C_{\text{f} ,1}^{(4)}C_{n\bm{\omega}_ m,0}^{(4)} \right) \ , \nonumber \\
&f_{c;n\bm{\omega}_ m}^{(5)} = f_{c;n\bm{\omega}_ m}^{(2)}\frac{4}{5} + \frac{(1-N)^2}{4(1+N)^2} + \frac{1-N}{5(1+N)^2} + \frac{1}{25(1+N)^2} \\
&  + \left( C_{\text{f},0}^{(3)}C_{n\bm{\omega}_ m,1}^{(3)} + C_{\text{f},1}^{(3)}C_{n\bm{\omega}_ m,0}^{(3)} \right)\frac{12}{7} + \left( C_{\text{f},0}^{(4)}C_{n\bm{\omega}_ m,1}^{(4)} + C_{\text{f},1}^{(4)}C_{n\bm{\omega}_ m,0}^{(4)} \right)\cdot2 \nonumber \\
& + \left( C_{\text{f},0}^{(5)}C_{n\bm{\omega}_ m,1}^{(5)} + C_{\text{f},1}^{(5)}C_{n\bm{\omega}_ m,0}^{(5)} \right)+ C_{\text{f},0}^{(3)}C_{n\bm{\omega}_ m,0}^{(3)} \left[\frac{1}{2}\frac{1-N}{1+N}+\frac{6}{7(1+N)}+\frac{18}{49(1-N^2)}\right] \;. \nonumber 
\end{align}
From the $z^3$ order constraint, we obtain
\begin{align}
\frac{C_{n\bm{\omega}_ m,1}^{(3)}}{C_{n\bm{\omega}_ m,0}^{(3)}} &= N^3 +3nN^2 -3N -\frac{6}{N+2} -3n +4 \; ,
\end{align}
which reproduces the $1/c$ order term in \eqref{wexpN}.
From the $z^4$ order constraint, we find
\begin{align}
\tiny
\begin{aligned}
&\frac{C_{n\bm{\omega}_ m,1}^{(4)}}{C_{n\bm{\omega}_ m,0}^{(4)}}=\frac{1}{40nm (N-1) (N+1)^2 (N+2) (N+3) (m-N) \left(5 m^2-5 m N+N^2+1\right)}\\
&\hspace{15pt}\left[4 \left(-25 nm^4 (N-1)^2 (N+1)^3 (n (N-8) (N+2) (N+3)-2 (N (2 N (N+8)+29)+6))\right) \right. \\
&\hspace{30pt}+50 nm^3 (N-1)^2 N (N+1)^3 (n (N-8) (N+2) (N+3)-2 (N (2 N (N+8)+29)+6))\\
&\hspace{30pt}-5 nm^2 (N-1)^2 (N+1)^3 \left(n (N+2) (N+3) \left(N^2 (5 N-52)-2\right)-2 (N (6 N (N (2 N (N+8)+29)+7)+17)+6)\right)\\
&\hspace{30pt}-10 nm (N-1)^2 N (N+1)^3 \left(n (N+2) (N+3) \left(6 N^2+1\right)+N (N (N (2 N (N+8)+29)+12)+17)+6\right)\\
&\hspace{30pt}\left.+(N+2)^2 (N+3)^2 (7-5 N)^2\right]\\
&+\frac{1}{40}\left[-40 N^3-290 N^2-60 N-\frac{3780}{N-3}+\frac{320}{N-2}-\frac{47}{N-1}+\frac{12}{(N-1)^2}+\frac{27}{N+1}+\frac{12}{(N+1)^2}+\frac{240}{N+2}+\frac{1440}{N+3}-1610\right] 
\, .
\end{aligned}
\normalsize 
\end{align}
In the same way, we can obtain the expression of $C_{n\bm{\omega}_ m,1}^{(5)} / C_{n\bm{\omega}_ m,0}^{(5)}$, but the result is too complicated to write down here.

\section{sl(3) generators}
\label{sl3}

One of the aims in this paper is to generalize the analysis in \cite{Hikida:2017ehf} by examining correlators with general light operators for $N=3$.
We parametrize the representation of the operator by Dynkin labels as $(\lambda_1,\lambda_2 ; 0)$.
For the generalization, we need the expressions of sl$(3)$ generators in terms of $X = (x,y,w)$ and the highest and lowest weight states denoted as $\langle \text{lw}| X \rangle$ and $\langle X | \text{hw} \rangle$, respectively. 
For the sl(3) generators, we use the expressions in \cite{DiFrancesco:1997nk} as
\begin{align}
&f_0^1 = - \frac{\partial}{\partial x} \, , \quad
f_0^2 = - \frac{\partial}{\partial y} - x \frac{\partial}{\partial w} \, , \quad
f_0^3 = -\frac{\partial}{\partial w} \, , \nonumber \\
&h_0^1 = - \lambda_2 + 2 x \frac{\partial}{\partial x} - y \frac{\partial}{\partial y} + w \frac{\partial}{\partial w} \, , \quad
h_0^2 = - \lambda_1 - x \frac{\partial}{\partial x} + 2 y \frac{\partial}{\partial y} + w \frac{\partial}{\partial w} \, , \\
&e_0^1 = - \lambda_2 x + w \frac{\partial}{\partial y} + x^2 \frac{\partial}{\partial x} + x w \frac{\partial}{\partial w} - x y \frac{\partial}{\partial y} \, , \quad
e_0^2 = - \lambda_1 x - w \frac{\partial}{\partial x} + y^2 \frac{\partial}{\partial y} \, , \nonumber \\
&e_0^3 = - ( \lambda_1 + \lambda_2 ) w + \lambda_1 x y + x w  \frac{\partial}{\partial x} + y w \frac{\partial}{\partial y} + w^2 \frac{\partial}{\partial w} - x y^2 \frac{\partial}{\partial y}\, . \nonumber
\end{align}
The general states are expressed by monomials of the form $x^r y^s w^t$.
The lowest weight state is represented as
$
\langle X | \text{lw} \rangle = 1 .
$
The state is annihilated by the lowering operators $f_0^a$ $(a=1,2,3)$,
and the eigenvalues of the Casimir generators $h_0^i$ $(i=1,2)$ are $( - \lambda_2 , - \lambda_1)$.
This implies that we should use (see also \cite{Fitzpatrick:2016mtp,Hikida:2017ehf})
\begin{align}
\langle \text{lw} | X\rangle = \delta (x) \delta (y) \delta (w) \, .
\label{Xlw}
\end{align}
The highest weight states are annihilated by $e^a_0$ $(a=1,2,3)$, and  the eigenvalues of $h_0^i$  are $( \lambda_1 , \lambda_2)$.  These conditions can be solved by
\begin{align}
\langle X | \text{hw} \rangle =
2^{\lambda_1 + \lambda_2} (x y - w)^{\lambda_1} w^{\lambda_2}\, ,
\label{Xhw}
\end{align}
where the overall factor is chosen such that the expression in \eqref{2ptN3tree} becomes simpler.

For the application to higher spin gravity, it is important to decompose the sl(3) generators in terms of embedded sl(2). 
Here, we choose
\begin{align}
V_{1}^{2} = - f_0^1 - f_0^2 \, , \quad V_0^{2} = h_0^1 + h_0^2 \, , \quad
V_{-1}^{2} =  2 ( e_0^1 + e_0^2 ) \, ,
\end{align}
and
\begin{align}
V_{2}^{3} =  f_0^3 \, , \quad
V_{1}^{3} = - \frac12 ( f_0^1 - f_0^2 ) \, , \quad V_0^{3} = \frac13 ( h_0^1 - h_0^2 ) \, , \quad
V_{-1}^{3} =   - e_0^1 + e_0^2 \, , \quad
V_{-2}^{3} = 4  e_0^3 \, .
\end{align}
We can show that the commutation relations
\begin{align}
\begin{aligned}
& [V_{m}^{2} , V_n^{2}] = (m-n) V_{m+n}^{2} \, , \quad
[V_{m}^{2} , V_n^{3}] = (2 m - n) V_{m +n}^{3} \, , \\
& [V_{m}^{3} , V_{n}^{3}] = - \frac{1}{12} (m - n) (2 m^2 + 2n^2 - m n - 8)  V_{m+n}^{2}
\end{aligned}
\end{align}
are satisfied.

For the manipulation of Wilson line operators, we need some algebraic computations.
In \eqref{fh} we may use
\begin{align}
e^{ z V_1^{2}} \langle X|  \text{lw} \rangle
= 2^{\lambda_1 + \lambda_2}
( (x + z) (y + z) -  (\tfrac12 z^2 + z x + w))^{\lambda_1}
( \tfrac12 z^2 + z x + w)^{\lambda_2} \, , \label{xshift}
\end{align}
where we have utilized the following formula:
\begin{align}
e^{ z V_1^{2}} f(X) = e^{ z \left( \frac{\partial}{\partial x} +  \frac{\partial}{\partial y} + x \frac{\partial}{\partial w} \right)} f(X)  =  e^{\left( \frac{z^2}{2} + z x \right) \frac{\partial}{\partial w} } e^{z  \frac{\partial}{\partial y}}e^{z  \frac{\partial}{\partial x} }  f(X) \, .
\label{BCH}
\end{align}
Similarly, we can compute
\begin{align}
\begin{aligned}
e^{z V_1^{2}} V_{-1}^{2} e^{ - z V_1^{2}}
&= - 2 \Biggl[ \lambda_2 x_{z} + \lambda_1 y_{z}  + (x_{z} y_{z} - w_{z}  - y_{z}^2 ) \frac{\partial}{\partial y}\\
&  \qquad \qquad + (w_{z}- x_{z}^2) \left( \frac{\partial}{\partial x} - z  \frac{\partial}{\partial w} \right) - x_{z} w_{z} \frac{\partial}{\partial w} \Biggr]  \, , \label{gshift} \\
e^{zV_1^{2}} V_{-2}^{3} e^{ - z V_1^{2}}
&=  - 4 \Biggl[ ( \lambda_1 + \lambda_2 ) w_{z} - \lambda_1 x_{z} y_{z}  - x_{z} w_{z}  \left( \frac{\partial}{\partial x} -z \frac{\partial}{\partial w} \right)  \\
&  \qquad \qquad + (x_{z} y_{z}^2 - y_{z} w_{z}) \frac{\partial}{\partial y} - w_{z}^2 \frac{\partial}{\partial w}  \Biggr] 
\end{aligned}
\end{align}
with
\begin{align}
x_{z} = x + z   \, , \quad y_{z} = y + z   \, , \quad w_{z} = w + z x + \tfrac12 z^2  \, ,
\label{gshift2}
\end{align}
which are used for \eqref{fh}.

\section{Simple example of general W$_3$ block}
\label{simple}

In the main context, we examined general blocks with the Wilson line method for $N=2$ up to the next leading order in $1/c$ and for $N=3$ at the leading order in $1/c$.
In this appendix, we examine the $1/c$ correction of the general W$_3$ block for \eqref{4pt1} with $N=3$ and $n=m=1$. For the simple case, it is not difficult to apply the technique in \cite{Besken:2016ooo} for algebraic calculations, even though our approach with the $X$-basis should be useful for more complicated examples with generic representations just as the leading order analysis in subsection \ref{largec}.

Similarly to the $N=2$ case, we start by examining the three point function
\begin{align}
\langle \mathcal{O}_{(\bar{\text{f}};0)} (z) \mathcal{O}_{(\text{f};0)} (0) \mathcal{O}_{(\text{adj};0)} (1) \rangle  \propto \frac{1}{(1 -z)^{ h_\text{adj}}z^{2 h_\text{f} - h_\text{adj}}} \label{3ptsimple}
\, ,
\end{align}
where adj represents the adjoint representation of su(3).
The conformal dimensions are
\begin{align}
h_\text{f} \equiv h(\text{f}; 0) = - 1 - \frac{32}{c} + \mathcal{O}(c^{-2}) \, , \quad
h_\text{adj} \equiv h(\text{adj}; 0) = - 2 - \frac{72}{c} + \mathcal{O}(c^{-2}) \, ,
\label{hexpsimple}
\end{align}
see \eqref{hexp}.
In the Wilson line network of \eqref{WLN} with $n=3$, we set $(z_1,z_2,z_3)=(z,0,1)$ and $(\Lambda_1,\Lambda_2,\Lambda_3) = (\bar{\text{f}} , \text{f} , \text{adj}  )$.
Putting $z_0=z_3 =1$, we compute the $1/c$ corrections from the diagrams in figure \ref{3ptWilson1}. 
For the leading order in $1/c$, we use
\begin{align}
\begin{aligned}
&q^{(i)}_j \equiv \langle e_j | e^{z_{0i} V_1^2} | e_1 \rangle_i  = \delta_j^1 - \sqrt{2} z_{0i} \delta_j^2 + z_{0i}^2 \delta^3_j \, , \\
&\bar{q}_{(i)}^j \equiv \langle \bar{e}_j | e^{z_{0i} V_1^2} | \bar{e}_3 \rangle_i  = \delta_j^3 + \sqrt{2} z_{0i} \delta_j^2 + z_{0i}^2 \delta^1_j 
\end{aligned}
\end{align}
by adopting the conventions in \cite{Castro:2011iw,Besken:2016ooo}.
For examples, $|e_j \rangle_i$ and $|\bar{e}_j \rangle_i$ denote the states in the fundamental and anti-fundamental representations, which are used to construct the general representation $\Lambda_i$.
As a singlet $\langle S |$, we first pick up the adjoint representation among $\Lambda_1 \otimes \Lambda_2$ and then construct a singlet with $\Lambda_3 = \text{adj}$.
This choice of singlet leads to 
\begin{align}
\left. \langle G_3 (\Lambda_i | z_i) \rangle \right|_{\mathcal{O}(c^0)} = \bar{q}_{(1)}^i q^{(2)}_j \bar{q}_{(3)}^j q^{(3)}_i  = (1-z)^2 \, ,
\end{align}
which reproduces the leading order term in \eqref{3ptsimple}.
For the next leading order in $1/c$,
we define
\begin{align}
&q^{(i;s)}_j (z')  \equiv  \frac{6}{c} \langle e_j |e^{ (z_0 - z') V_1^2} V^{s}_{-s+1} e^{ (z' - z_0) V_1^2} e^{z_{0i} V_1^2} | e_1 \rangle_i  \, , \nonumber \\
&\bar{q}_{(i;s)}^j (z')\equiv  \frac{6}{c} \langle \bar{e}_j | e^{ (z_0 - z') V_1^2} V^{s}_{-s+1} e^{ (z' - z_0) V_1^2} e^{z_{0i} V_1^2} | \bar{e}_3 \rangle_i \, , \\
&q^{(i;s,s)}_j (z',z'')  \equiv \left(  \frac{6}{c} \right)^2 \langle e_j |e^{ (z_0 - z') V_1^2} V^{s}_{-s+1} e^{ (z' - z_0) V_1^2}e^{ (z_0 - z'') V_1^2} V^{s}_{-s+1} e^{ (z'' - z_0) V_1^2}e^{z_{0i} V_1^2} | e_1 \rangle_i  \, , \nonumber \\
&\bar{q}_{(i;s,s)}^j (z',z'')\equiv \left(  \frac{6}{c} \right)^2 \langle \bar{e}_j | e^{ (z_0 - z') V_1^2} V^{s}_{-s+1} e^{ (z' - z_0) V_1^2}e^{ (z_0 - z'') V_1^2} V^{s}_{-s+1} e^{ (z'' - z_0) V_1^2} e^{z_{0i} V_1^2} | \bar{e}_3 \rangle_i \, . \nonumber
\end{align}
Just like the leading order, contributions at order $1/c$ come from the integrals as
\begin{align}
\begin{aligned}
& \int_z^1 d z_2 \int_z^{z_2} d z_1 \,  \bar{q}_{(1;s,s)}^i (z_2,z_1)q^{(2)}_j  \bar{q}_{(3)}^j q^{(3)}_i  \langle J^{(s)}(z_2) J^{(s)} (z_1) \rangle \, , \\
&\int_1^0 d z_2 \int_1^{z_2} d z_1 \, \bar{q}_{(1)}^i q^{(2;s,s)}_j  (z_1,z_2)  \bar{q}_{(3)}^j q^{(3)}_i  \langle J^{(s)}(z_2) J^{(s)} (z_1) \rangle\, , \\
& \int_z^1 d z_2 \int_1^{0} d z_1  \, \bar{q}_{(1;s)}^i (z_2)q^{(2;s)}_j (z_1)   \bar{q}_{(3)}^j q^{(3)}_i  \langle J^{(s)}(z_2) J^{(s)} (z_1) \rangle
\end{aligned}
\end{align}
with $s=2,3$ and \eqref{2ptcurrentr} with \eqref{B2B3}. The sum of them is computed as
\begin{align}
\left. \langle G_3 (\Lambda_i | z_i) \rangle \right|_{\mathcal{O}(c^{-1})} = \frac{1}{c} (1 - z)^2 (72 \log (1 - z )- 8 \log (z))
\end{align}
for the $\epsilon$-independent part. With the shifts of conformal weight in \eqref{hexpsimple}, we can see that this is the result expected from the conformal symmetry as in \eqref{3ptsimple}.

The four point block for \eqref{4pt1} with $N=3$ and $n=m=1$ can be examined in a similar manner.
Since we have already examined the identity block in subsection \ref{VacuumCB}, we focus on the W$_3$ block with the exchange of adjoint operator in \eqref{coulomb}.
The expression in $1/c$ expansion is
\begin{align}
\begin{aligned}
&(1-z)^2-\frac{z^2}{3} + \frac{1}{c} \Biggl[ \left((1-z)^2-\frac{z^2}{3}\right) (64 \log (1-z)-8 \log (z))\\
& \qquad+24 z (1-z)^2  -8 (1-z)^2 \sum _{n=2}^{\infty } (n-1) \left(3 H_{n-2}+\frac{1}{n}-3\right) z^n\Biggr] + \mathcal{O}(c^{-2}) \, .
\end{aligned}
\label{4pt11/c}
\end{align}
In the Wilson line network \eqref{WLN} with $n=4$, we set $(z_1,z_2,z_3,z_4)=(z,0,\infty,1)$ and  $(\Lambda_1,\Lambda_2,\Lambda_3 , \Lambda_4) = (\bar{\text{f}}, \text{f},\bar{\text{f}}, \text{f} )$.  
We may fix $z_0 = z_4 =1$ as in figure \ref{3ptWilson2}.
For the singlet $\langle S |$, we first pick up the adjoint representations in $\Lambda_1 \otimes \Lambda_2$ and $\Lambda_3 \otimes \Lambda_4$ and then construct a singlet from the product of the two adjoint representations. 
At the leading order in $1/c$, we find
\begin{align}
\left. \langle G_4 (\Lambda_i | z_i) \rangle \right|_{\mathcal{O}(c^0)} = \bar{q}^i_{(1)} q^{(2)}_j \bar{q}^j_{(3)} q^{(4)}_i - \frac13 \bar{q}^i_{(1)} q^{(2)}_i \bar{q}^j_{(3)} q^{(4)}_j = (1-z)^2 - \frac{z^2}{3} \, ,
\end{align}
which reproduces the leading order term in \eqref{4pt11/c}. Here, we should notice that the products $\bar{q}^i_{(1)} q^{(2)}_j \bar{q}^j_{(3)} q^{(4)}_i $ and $\bar{q}^i_{(1)} q^{(2)}_i \bar{q}^j_{(3)} q^{(4)}_j$ are the same as the identity blocks for the $z \to 1$ and $z \to 0$ channels, respectively. Since the structure of contractions among  indices does not change even at the next leading orders in $1/c$, the computation for the Wilson line network reduces to 
\begin{align}
 \langle G_4 (\Lambda_i | z_i) \rangle  = \langle W_\text{f} (z;1) W_\text{f} (\infty;0) \rangle - \frac13  \langle W_\text{f} (z;0) W_\text{f} (\infty;1) \rangle 
\end{align}
at least up to order $1/c$.
Using \eqref{VBself} and \eqref{G123}, the $1/c$ order terms of the quantity are obtained as
\begin{align}
\begin{aligned}
&\left. \langle G_4 (\Lambda_i | z_i) \rangle \right|_{\mathcal{O}(c^{-1})}  =  \frac1c \biggl\{ \biggl[ 64 (1-z)^2 \log (1-z) + 2 (1-z)^4 \, _2F_1(2,2;4;1-z) \\
& \qquad \qquad -\frac{2}{15} (1-z)^5 \, _2F_1(3,3;6;1-z) \biggr] -\frac{1}{3} \biggl[ 64 z^2 \log (z) + 2 z^4 \, _2F_1(2,2;4;z) \\
&\qquad \qquad -\frac{2}{15} z^5 \, _2F_1(3,3;6;z) \biggr] \biggr\}\, . 
\end{aligned}
\label{4pt1Wilson}
\end{align}
The difference between the $1/c$ order terms in \eqref{4pt11/c} and \eqref{4pt1Wilson} is proportional to the leading order terms in \eqref{4pt11/c}, which can be removed by changing the overall factor of the conformal block. Thus, we conclude that the Wilson line computation reproduces the CFT one in \eqref{coulomb} up to the $1/c$ order for the simple example.

\section{Conformal transformation of correlators}
\label{Shwarz}

In appendix A of \cite{Besken:2016ooo}, it was analyzed how global conformal transformations act on  correlators from Wilson line networks at the leading order in $1/c$. Here, we would like to extend the analysis to local conformal transformations.

We start from the gauge field 
\begin{align}
a(z) = V_1^2 + \frac{6}{c} T(z) V_{-1}^2 
\label{az2}
\end{align}
as in \eqref{az} with $N=2$ and $ T (z) = J^{(2)} (z)$. 
For the current analysis, we neglect the shift of coupling $c_2$ in \eqref{ren}.
We consider the su(2) gauge transformation as
\begin{align}
a' (z) = L(z) a(z) L(z)^{-1} + L(z) \frac{d}{dz} L(z)^{-1} 
\end{align}
with
\begin{align}
L(z) = e^{c_{-1} (z) V_{-1}^2} e^{2 \log c_0 (z)  V_{0}^2} e^{c_1 (z) V_{1}^2} \, .
\end{align}
We require that the $a'(z)$ is of the form \eqref{az2}, i.e., the coefficient of $V_0^2$ to be zero. 
We consider the following solution to this condition:
\begin{align}
c_1 (z) = 0 \, , \quad c_{-1} (z) = - c_0 (z) c_0' (z) \, .
\end{align}
This solution reduces to (A.6) of \cite{Besken:2016ooo} for global transformation induced by $c_0 (z) = c z + d$.

Under the gauge transformation, the open Wilson line becomes
\begin{align}
 P \exp \left\{ \int_{z_i}^{z_f} dz  \left[ \frac{1}{c_0(z)^2}V_1^2  + c_0 (z)^2 \left( \frac{6}{c} T (z) + \frac{c_0 ''(z)}{c_0 (z) } \right) V_{-1}^2 \right] \right\}\, .
\end{align}
In order to set the coefficient of $V_{1}^2$ to be 1, we change the coordinate $z$ to $w$ with
\begin{align}
c_0 (z)  = \left(  \frac{dz}{dw}\right)^{\frac12} \, .
\end{align}
The open Wilson line is now rewritten as
\begin{align}
P \exp \left( \int_{w_i}^{w_f} d w \,   a(w)  \right) \, , \quad
a(w) = V_1^2  +  \frac{6}{c} T (w)  V_{-1}^2 
\end{align}
with
\begin{align}
T(w) = \left( \frac{dw}{dz} \right)^{-2} \left( T(z) - \frac{c}{12} S (w,z) \right)  \, .
\label{Tw}
\end{align}
Here we have defined the Schwarzian derivative as
\begin{align}
S(w,z) = \frac{d ^3 w / dz^3}{(dw /dz)} - \frac{3}{2} \left( \frac{d^2 w / dz^2}{dw/dz}\right)^2 \, .
\label{sd}
\end{align}

Let us consider how a two point function changes under the transformation. We first adopt the regularization in \cite{Fitzpatrick:2016mtp}, where we take the normal ordered prescription of $T(z)$ in the same open Wilson line and replace the quantum number $h_0 = - j$ by the exact conformal weight $h_j$. 
Then, we find
\begin{align}
\begin{aligned}
\langle \text{lw} | P \exp \left(\int_{z_i}^{z_f} dz \, a(z) \right) | \text{hw} \rangle 
= \langle \text{lw} | L^{-1} (z_f) P \exp \left(\int_{z_i}^{z_f} dz \, a ' (z) \right) L (z_i) | \text{hw} \rangle  \\
= \left( \frac{dz_f}{d w_f} \right)^{- h_j} \left( \frac{dz_i}{d w_i} \right)^{- h_j}
 \langle \text{lw} | P \exp \left(\int_{w_i}^{w_f} d w \, a(w) \right) | \text{hw} \rangle \, ,
\end{aligned}
\label{trans1}
\end{align}
see (A.10) of \cite{Besken:2016ooo}. This is consistent with the local conformal transformation of two point function as
\begin{align}
\langle \mathcal{O}_j (z_f) \mathcal{O}_j (z_i) \rangle =
 \left( \frac{dz_f}{d w_f} \right)^{- h_j} \left( \frac{dz_i}{d w_i} \right)^{- h_j}
 \langle \mathcal{O}_j (w_f) \mathcal{O}_j (w_i) \rangle \, .
 \label{trans2}
\end{align}
With our prescription, we regularized the open Wilson line for the two point function when the gauge field corresponds to the AdS$_3$ background and the coordinate $z$ is the planar one. 
For other cases, we offer a way to regularize it such that \eqref{trans2} is satisfied, and this is the prescription adopted in section \ref{HLcorrelator}.

%\bibliographystyle{JHEP}
%\bibliography{HSSG}

\providecommand{\href}[2]{#2}\begingroup\raggedright\endgroup

\end{document}